\documentclass[journal]{IEEEtran}

\usepackage{cite}
\usepackage{amsmath,amssymb}
\usepackage{graphicx}
\usepackage{booktabs}
\usepackage{tabularx}
\usepackage{array}
\usepackage{enumitem}
\usepackage[table]{xcolor}
\usepackage{url}
\usepackage{xurl}
\usepackage{microtype}

\usepackage{booktabs,tabularx,array,ragged2e}
\usepackage{xcolor}
\usepackage{colortbl}

\definecolor{RoadHead}{HTML}{E9F0F6}
\definecolor{RoadInk}{HTML}{243E52}
\definecolor{RoadRule}{HTML}{C5D1DA}

\newcolumntype{R}[1]{>{\RaggedRight\arraybackslash}p{#1}}
\newcolumntype{Z}[1]{>{\hsize=#1\hsize\linewidth=\hsize
  \RaggedRight\arraybackslash}X}

\renewcommand{\arraystretch}{1.08}
\newcolumntype{L}[1]{>{\raggedright\arraybackslash}p{#1}}
\newcolumntype{C}[1]{>{\centering\arraybackslash}p{#1}}
\newcolumntype{Y}{>{\raggedright\arraybackslash}X}
\definecolor{TableHead}{gray}{0.92}
\definecolor{TableGroup}{gray}{0.965}
\definecolor{TableAccent}{gray}{0.94}

\title{Voice Agents: From Auditory Intelligence to Auditory Agency}

\author{Kele Xu,~\IEEEmembership{Senior Member,~IEEE}, Qisheng Xu, Xinwang Liu,~\IEEEmembership{Senior Member,~IEEE}, Huaimin Wang
\thanks{Kele Xu is with the School of Computer Science, National University of
Defense Technology, Changsha 410073, China, and with the State Key Laboratory
of Complex \& Critical Software Environment, Changsha 410073, China (e-mail:
kelele.xu@gmail.com; ORCID: 0000-0002-7136-1898).}%
}

\begin{document}

\maketitle

% Override the IEEEtran default running heads for this venue.
\markboth{Proceedings of the IEEE}{K. Xu: From Auditory Intelligence to Auditory Agency}

\begin{abstract}
Voice systems increasingly combine streaming speech, persistent state, reasoning, tool use, and embodied control. Once these functions share a runtime, spoken input can change digital or physical state, and recognition or response quality alone no longer characterizes the system. This article studies the interfaces through which spoken evidence becomes consequential. \emph{Auditory intelligence} denotes the capabilities required to perceive, understand, coordinate, and generate spoken interaction; \emph{auditory agency} denotes the causal path from that interaction to stateful decisions and observable consequences. The \emph{commitment boundary} marks where revisable auditory evidence becomes persistent state, a plan, or an executable request. \emph{Agency closure} then follows an authorized action through observation, verification, recovery or compensation, and communication of the outcome. Evidence from full-duplex interaction, reasoning, memory, tools, embodiment, and human control shows a consistent systems-level mismatch: integration is advancing faster than end-to-end observation of what the integrated system actually does. The resulting design problem is to preserve a trace from spoken evidence to consequence while keeping commitment, execution, and verification attributable.
\end{abstract}

\begin{IEEEkeywords}
Voice agents, Spoken dialogue systems, Speech foundation models, Auditory agency, Embodied agents
\end{IEEEkeywords}

\section{Introduction}
\label{sec:intro}

Recent advances have extended voice systems well beyond speech recognition and response generation. Some systems can listen while speaking, while others maintain information across sessions, invoke external tools, or connect spoken instructions to embodied controllers \cite{PAPER_1473A84B8A9D,PAPER_07DE94B0DB13,PAPER_B421DE5BB9A8,EXTERNAL_ELLSA}. Much of this progress has been evaluated through individual capabilities, with less attention to how they operate together. The difficulty becomes apparent when a spoken request is still unfolding, or is being corrected, while the system has already begun preparing an action. Even accurate recognition and fluent responses cannot establish whether the request was correctly interpreted, whether the action was authorized, or whether the resulting external state matches the user's intent.

This distinction motivates our use of two related concepts. We use \emph{auditory intelligence} to describe the perceptual, linguistic, and conversational abilities involved in spoken interaction \cite{REVIEW_ANCHOR_NMI2026}. \emph{Auditory agency}, by comparison, concerns the extent to which these abilities are coupled to stateful decisions and observable consequences. Existing systems approach this problem from different directions. Moshi, SALM-Duplex, and Freeze-Omni investigate speech overlap and interaction timing \cite{PAPER_1473A84B8A9D,PAPER_FAE46C323E2F,PAPER_65418243EE76}. AURA and AnovaX introduce planning or execution interfaces \cite{PAPER_B421DE5BB9A8,PAPER_40A1EF47A5D0}, while embodied systems connect spoken instructions to navigation and manipulation \cite{PAPER_A40BBDA83DC5,EXTERNAL_ELLSA}. These developments raise a common systems question: how should an agent turn provisional spoken input into an authorized commitment, and how can it determine whether the intended action has actually been carried out?

\subsection{Why Voice Agents Have Become a Systems Problem}
\label{sec:intro-why}

Speech-native foundation models have brought acoustic processing and language modeling into closer integration. AudioPaLM demonstrated joint modeling of speech and language within a shared model \cite{PAPER_190E84406795}, while SALMONN-omni and MinMo have advanced streaming and unified voice interaction \cite{PAPER_C5043A3C2B57,PAPER_F09AF490D578}. Alongside these developments, semantic endpointing and incremental understanding allow systems to interpret meaningful parts of an utterance before the speaker has finished \cite{PAPER_C69F6F77FA7A,PAPER_CB8E8D2C7474}. Moshi, SALM-Duplex, and Freeze-Omni further support overlapping speech input and output, moving beyond the conventional alternation between listening and speaking \cite{PAPER_1473A84B8A9D,PAPER_FAE46C323E2F,PAPER_65418243EE76}. A voice system can thus begin interpreting and responding to a request while the speaker is still formulating or revising it.

The implications extend beyond conversational timing. Incremental semantic information can initiate reasoning before a complete transcript is available \cite{PAPER_CB6C6E4E2062,PAPER_13438B5C64C7}. Memory mechanisms can carry user preferences and interaction histories across sessions, with identity-aware designs seeking to keep this information associated with the appropriate user \cite{PAPER_FFAF306C931D,PAPER_07DE94B0DB13,PAPER_4AC64B8C94AB}. Speech-to-function models and synchronized action streams provide interfaces between spoken requests and structured commands \cite{PAPER_766B59A1727C,PAPER_09240145263B}, while multimodal and embodied systems connect speech to navigation, manipulation, and situated assistance \cite{PAPER_A40BBDA83DC5,PAPER_E9854767F763,PAPER_A8347D87D2F8,EXTERNAL_ELLSA}. These capabilities bring interpretation, reasoning, memory, and action into increasingly close contact. An interpretation formed early in an utterance may now influence decisions or downstream behavior before the speaker's intention is fully established.

Managing such dependencies is not entirely new. Classical spoken dialogue systems already combined dialogue state, policy, service access, and speech synthesis, typically through modules with identifiable interfaces and responsibilities \cite{PAPER_E17F117D3387,PAPER_03CE3B0B842C,PAPER_86E0C27FBC57}. Foundation-model architectures allow these functions to interact more closely, often without equally explicit boundaries between perception, dialogue management, and planning. This integration can make spoken interaction more fluid, but it also makes responsibility for intermediate decisions harder to locate. A system must still determine when a provisional interpretation is sufficiently established to support an authorized action, which component is responsible for that decision, and how subsequent revisions should be handled.

\subsection{Auditory Agency, Commitment, and Closure}
\label{sec:intro-thesis}
\begin{figure*}[htbp]
\centering
\includegraphics[width=\textwidth]{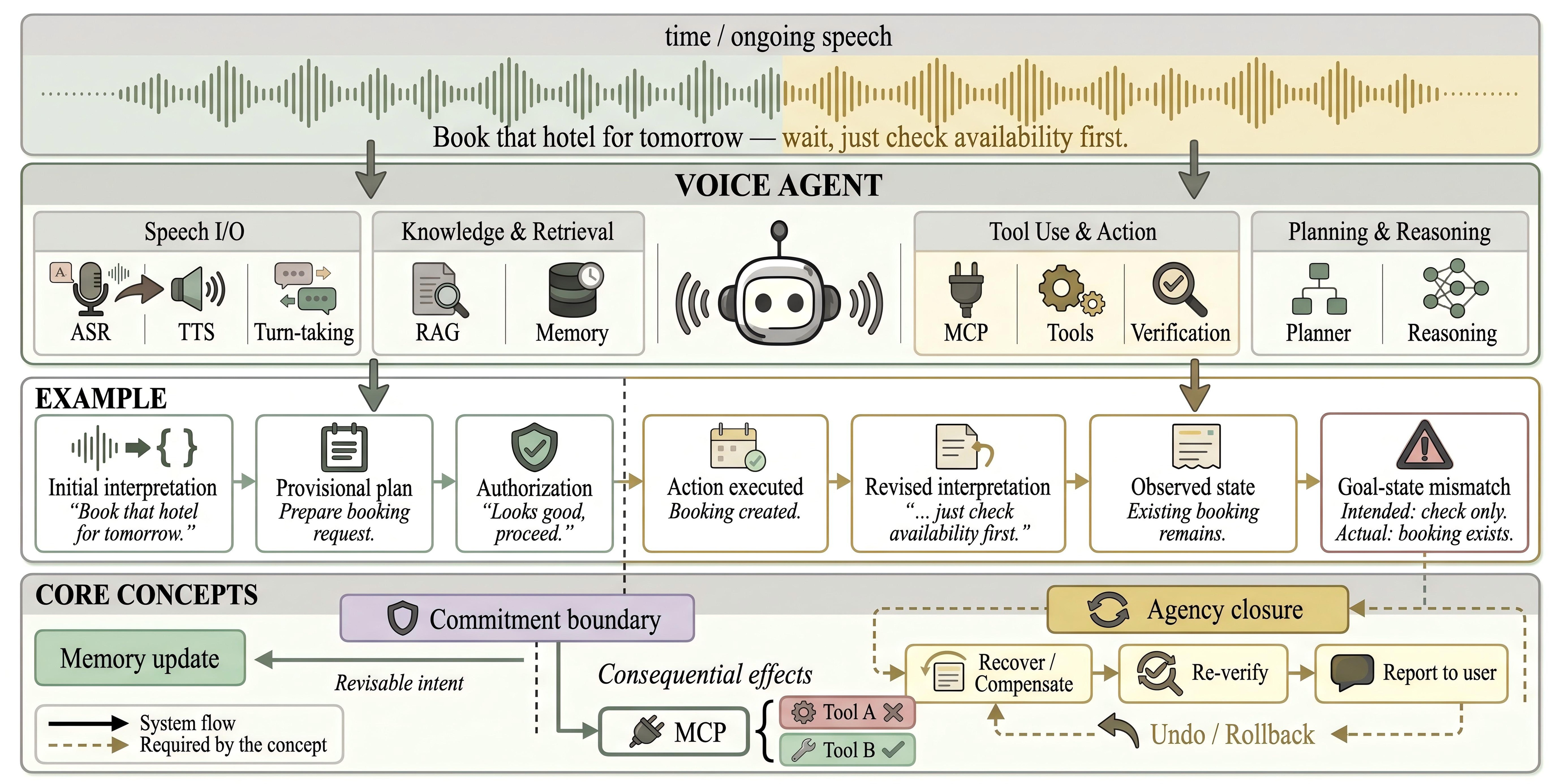}
\caption{From auditory intelligence to auditory agency. The figure illustrates how partial spoken input can trigger premature actions that conflict with subsequent user corrections. It highlights the commitment boundary between provisional interpretation and downstream state or action, and introduces agency closure as the process of verifying consequences, handling mismatches, and reporting outcomes to the user. Solid arrows depict the illustrative failure scenario, while dashed arrows indicate the required control and recovery paths.}
\label{fig:intelligence-agency}
\end{figure*}

Interaction intelligence and agentic intelligence are separate analytical dimensions. Interaction intelligence concerns participation in spoken exchange-perception, source and addressee attribution, incremental understanding, turn coordination, speech generation, and repair. Agentic intelligence concerns goal-directed state that persists beyond the immediate utterance-reasoning, planning, memory, authorization, action, observation, verification, and recovery. Neither dimension is a maturity scale. The systems problem appears where the two meet and conversational evidence begins to carry consequences.

We call this point the \emph{commitment boundary}. Before commitment, hypotheses about words, intent, addressee, or goal can remain provisional. After commitment, an interpretation has been written into persistent state, used to instantiate a plan, or converted into a request that is eligible for execution. In a streaming system, the acoustic endpoint and the commitment decision need not coincide. The former asks whether speech appears to have stopped; the latter asks whether enough evidence exists for downstream state or action to depend on the current interpretation. Keeping the two events separate makes premature actions easier to identify and, in principle, easier to retract.

The second concept is \emph{agency closure}. An intended action is authorized and executed; the resulting external state is then observed and checked against the intended state; a mismatch triggers recovery or compensation; and the outcome is communicated back to the user. This definition keeps several commonly conflated events apart. Tool selection is not tool execution. Execution is not evidence that the intended state changed. Observation of a state is not yet verification that the user's goal was met. The establishment and repair of shared understanding have long been studied as grounding in communication and in multimodal task-oriented collaboration~\cite{clark1991grounding,dillenbourg1996grounding}.

Memory, authorization, and recovery fit naturally into this account. Persistent memory carries identity, provenance, and temporal scope. Authorization determines which otherwise valid plans may cross into execution. Recovery determines what happens after a failed or partial state transition. Later sections return to four concrete questions: what was committed, who or what owned the decision, what evidence established the consequence, and what happened when intended and observed state disagreed.

\subsection{Scope and Contributions}
\label{sec:intro-scope}

A \emph{voice agent} is used here to mean a system that receives speech or richer auditory input, maintains interaction or task state, selects a response or action, and produces speech or another observable consequence. The organizing unit is not a product category or a model family, but the interface through which auditory evidence becomes persistent state, a decision, an action, or an outcome. Classical dialogue systems, speech-language models, duplex interaction, memory, tool use, embodied control, and human-centered evaluation are therefore discussed where they illuminate one of those interfaces.

The analysis follows the path from a spoken request to its consequence. The commitment boundary and agency closure provide the two reference points along that path. Architectures are compared by who owns state, decisions, execution, and verification; evaluation results are read according to the endpoint they actually observe. This keeps spoken uncertainty, authority, persistent state, and external consequence visible without forcing them onto a single maturity scale.

The discussion proceeds from architecture (Section~\ref{sec:evolution}) and interaction intelligence (Section~\ref{sec:interaction}) to agentic intelligence (Section~\ref{sec:agentic}), human-centered control (Section~\ref{sec:human}), closed-loop evaluation (Section~\ref{sec:evaluation}), and the research agenda that follows from these interfaces (Section~\ref{sec:roadmap}).
\section{Evolution and Architecture of Voice Agents}
\label{sec:evolution}

Whether a voice agent can revise an early interpretation or recover from a failed action depends in part on its architecture. Modular systems expose uncertainty and responsibility at interfaces; integrated models can make those boundaries harder to locate. Speech generation, streaming understanding, turn control, retrieval, reasoning, and safety evolved along partly separate lines and are now being brought into common runtimes. As these functions begin to share state, the placement of decisions, authorization, and recovery becomes a central design concern.

\subsection{From Spoken Dialogue Systems to Foundation Models}
\label{sec:evolution-history}

\textbf{Recognition plus policy.} Early spoken dialogue systems organized recognition, dialogue state, policy selection, back-end access, and synthesis into identifiable stages. Probabilistic dialogue management treated interaction as decision-making under partial observability, representing uncertainty through belief states and optimizing policies against explicit rewards~\cite{young2013pomdp}. Plato exemplified the modular approach by exposing dialogue state, policy, and back-end services as replaceable components~\cite{PAPER_E17F117D3387}. State tracking also became more attentive to uncertainty in the speech signal. Word confusion networks carried alternative recognition hypotheses into the state estimate~\cite{PAPER_03CE3B0B842C}, while temporally expressive networks and online reinforcement learning explored joint optimization of belief tracking and policy~\cite{PAPER_86E0C27FBC57,PAPER_EC2365EC8FAE,PAPER_A2DB3D2CB7CB}. These systems established much of the vocabulary of state, policy, and reward still used in voice agents. Their explicit interfaces also made intermediate decisions comparatively easy to inspect, although evaluation generally remained tied to bounded domains and task-specific goals.

\textbf{Neural and speech-native modeling.} Neural modeling changed what could pass through those interfaces. Advances in speech synthesis made generated speech more suitable for conversational use~\cite{PAPER_79AAF689FC20}, while speech-to-speech translation increasingly avoided a text-only intermediate representation. Translatotron 2 addressed voice preservation during translation~\cite{PAPER_A224F6D9B6D8}; later approaches used learned discrete units~\cite{PAPER_BE36882FCC21}, two-pass decoding~\cite{PAPER_A729F1BBE19F}, and monolingual training data~\cite{PAPER_B591F4760692}, before scaling to multilingual settings in SeamlessM4T~\cite{PAPER_79C802D98E40}. AudioPaLM brought speech input and generation into a shared language-model formulation~\cite{PAPER_190E84406795}. SpeechGPT extended instruction following to discrete speech representations~\cite{zhang2023speechgpt}, and SpiRit-LM interleaved written and spoken tokens, adding pitch and style units in its expressive variant~\cite{nguyen2025spirit}. Two-channel generative dialogue models also showed that interactive speech could be modeled without a separately specified turn-taking module~\cite{PAPER_1305D1F3BECF}. These developments broadened the acoustic and linguistic capabilities of spoken systems. They did not, on their own, settle how a system should identify the addressee, maintain user-specific state, or withdraw an action after the speaker changes course.

\textbf{Streaming and temporal control.} A related line of research addressed the fact that speech arrives incrementally. Early incremental dialogue architectures represented partial interpretations as linked units that could be added, revised, or committed, allowing corrections to propagate through dependent modules~\cite{schlangen2011general}. Incremental generation brought a similar principle to the output side: a system could begin speaking before planning its entire response~\cite{skantze2010towards}. Streaming spoken language understanding subsequently produced intent and slot predictions from incomplete audio~\cite{PAPER_C69F6F77FA7A}, and response-timing models examined when a system should begin speaking as well as what it should say~\cite{PAPER_DE1FE99D0F91,PAPER_0BE6D6719412}. Work combining TurnGPT with voice activity projection connected learned timing predictions to conversational robots, including preparation, turn-taking, and interruption handling~\cite{skantze2025applying}. Endpointing also moved beyond silence detection. Phoenix-VAD uses semantic information to judge utterance completion~\cite{PAPER_CB8E8D2C7474}; adaptive detection addresses false interruptions caused by residual playback~\cite{PAPER_D9E5B1683133}; and fused models distinguish pauses, backchannels, interruptions, and continuations~\cite{PAPER_11D6F89D815C}. Device-directedness poses a separate question: whether the speech was addressed to the assistant in the first place~\cite{PAPER_40CE2E5F8957}. Together, these mechanisms make earlier responses possible, but they also increase the importance of revision. A late correction may arrive after reasoning has started or an operation has been prepared.

\textbf{Speech-language models and duplex dialogue.} Recent speech-language models have brought acoustic representation, language modeling, and conversational control into closer contact~\cite{PAPER_C5043A3C2B57,PAPER_F09AF490D578}. Duplex architectures go further by allowing listening and speaking to overlap. Moshi models parallel user and system codec-token streams alongside an internal text stream, with a reported theoretical latency of approximately 160\,ms~\cite{PAPER_1473A84B8A9D}. SALM-Duplex learns the decision to listen or speak~\cite{PAPER_FAE46C323E2F}, whereas Freeze-Omni adds duplex interaction to a frozen language model~\cite{PAPER_65418243EE76}. GLM-4-Voice explores end-to-end spoken dialogue~\cite{PAPER_42139A057A67}, and related work extends speech-language modeling to longer contexts~\cite{PAPER_30C552706E67}. Other designs retain distinct components: LLaMA-Omni 2 pairs a language-model backbone with a speech encoder and streaming autoregressive decoder~\cite{fang2025llama}; SyncLLM introduces temporal information to align generation with real-world time~\cite{liu2025sync}; and OmniFlatten uses flattened multimodal sequences and staged training to handle overlap, backchannels, and interruptions~\cite{zhang2025omniflatten}. Duplex control can also remain a pluggable module~\cite{PAPER_30744A6D4C42}, while structured-action models introduce a separate stream for planning and tool calls~\cite{PAPER_09240145263B}. Evaluation has followed these changes, extending from turn-taking and overlap handling~\cite{PAPER_760BC4ACD4D7} to automated multi-turn tests~\cite{PAPER_B8CB31F1F27C}, disfluent human recordings with mock tool APIs~\cite{PAPER_9AD353416653}, interruption-rich dialogue generation~\cite{PAPER_4F97D68B6C26}, and recovery after interruption~\cite{PAPER_2D9099D449E2}. The resulting benchmarks reveal different aspects of conversational timing; none alone establishes whether an external action was appropriate or completed.

\textbf{Reasoning, tools, memory, and grounding.} As speech interfaces became more responsive, researchers began connecting them to longer task sequences. ReAct, developed in the broader language-agent literature, interleaved reasoning, actions, and observations and provided an influential precedent for tool-oriented control loops~\cite{yao2022react}. Voice-oriented work has coordinated specialized audio models through a language-model controller~\cite{PAPER_E3E2EEA2C7C7}, initiated reasoning from incremental semantic events~\cite{PAPER_13438B5C64C7}, and applied chain-of-thought training to cascaded and streaming duplex dialogue~\cite{PAPER_CB6C6E4E2062,PAPER_08AED7C778E1}. Speech-to-function benchmarks test the conversion of noisy or disfluent requests into structured calls~\cite{PAPER_766B59A1727C,PAPER_2423A4A10719}. Agent frameworks make orchestration and typed execution explicit~\cite{PAPER_B421DE5BB9A8,PAPER_40A1EF47A5D0}, while grounded benchmarks compare a database end state against a specified goal~\cite{PAPER_535B4C25EDC0}. Domain-oriented systems illustrate the practical requirements of form filling, conversational search, and enterprise support~\cite{PAPER_3FD26CDF1D8B,PAPER_838318340492,PAPER_0BB1928924AB}. Beyond speech, $\tau$-bench evaluates tool-using agents with domain policies, API tools, simulated users, and repeated trials, measuring completion against an annotated final database state~\cite{yao2024tau}. Such evaluations bring the consequences of tool use into view, although simulated calls and verified changes in an external service remain different forms of evidence.

Recent multimodal releases illustrate how these functions are converging at the model level. Xiaomi's MiMo documentation describes native audio understanding and agent execution within the same multimodal model family~\cite{MIMO_V25_RELEASE_2026}. The Qwen3.5-Omni technical report likewise presents a shared foundation for audio-visual understanding, streaming speech generation, and native tool interfaces, including WebSearch and FunctionCall~\cite{team2026qwen3}. These are descriptions of model and platform capabilities, not independent evidence that external actions have been executed and verified. They nevertheless make the ownership of timing, memory, tool use, and outcome checking more pressing.

\textbf{Persistence, embodiment, and governance.} Another set of developments concerns what survives a turn and what happens outside the dialogue itself. Retrieval-augmented systems bring external knowledge into conversation~\cite{PAPER_76F78E88F459,PAPER_7C61510F1269,PAPER_45B6406FAE17}; memory mechanisms extract and reuse preferences across sessions~\cite{PAPER_FFAF306C931D,PAPER_07DE94B0DB13}; and long conversational audio exposes measurable degradation~\cite{PAPER_AA174144A864}. LongMemEval, although text-based, tests extraction, multi-session and temporal reasoning, knowledge updates, and abstention, offering useful evaluation distinctions for persistent state~\cite{wu2024longmemeval}. The same persistence raises security and privacy concerns. Speech-specific attacks and defenses~\cite{PAPER_AF925F6B6A7D,PAPER_FFE347059994}, benchmarks of interactional privacy~\cite{PAPER_49C7AB780EBD}, and governance frameworks for tool-using voice assistants~\cite{PAPER_BA00BDF2544F} examine different parts of this problem. Physical grounding adds another constraint. SayCan, from language-guided robotics, combines language-model knowledge with skill value functions to select actions feasible for a robot and its environment~\cite{ahn2022can}. Studies involving driving simulators, social robots, and clinical assessment show why such interfaces matter when spoken decisions affect situated tasks~\cite{PAPER_A40BBDA83DC5,PAPER_6688E5A9A34D,PAPER_D0815A91F238}.

The historical strands remain relevant to present-day systems. Translation quality, expressive synthesis, low-resource coverage, and safety still shape the interaction on which reasoning and action depend. Their evaluation measures, however, describe different endpoints. Tool-selection F-measure, turn-state classification accuracy, and task completion under realistic interaction conditions cannot be read as interchangeable measures of agency. Architectural comparison therefore needs to ask what each system makes observable, from speech behavior to executed actions and their consequences.

\subsection{Architectural Families}
\label{sec:evolution-families}
% ---------------------------------------------------------------------------
% Figure 2: parallel evolution of interaction intelligence and agentic
% intelligence, and the four-layer technology stack for voice agents.
% The artwork is an image file, so it is replaced by uploading a new PNG with
% the same name into figures/images/ in Overleaf.
% Source artwork: figures/images/fig2_parallel_evolution_stack.png
% (1536 x 926 px).
% The bitmap caption of the supplied artwork was trimmed; the caption below is
% typeset by LaTeX so that the figure keeps its number and cross-references.
% ---------------------------------------------------------------------------
\begin{figure*}[tp]
\centering
\includegraphics[width=\textwidth]{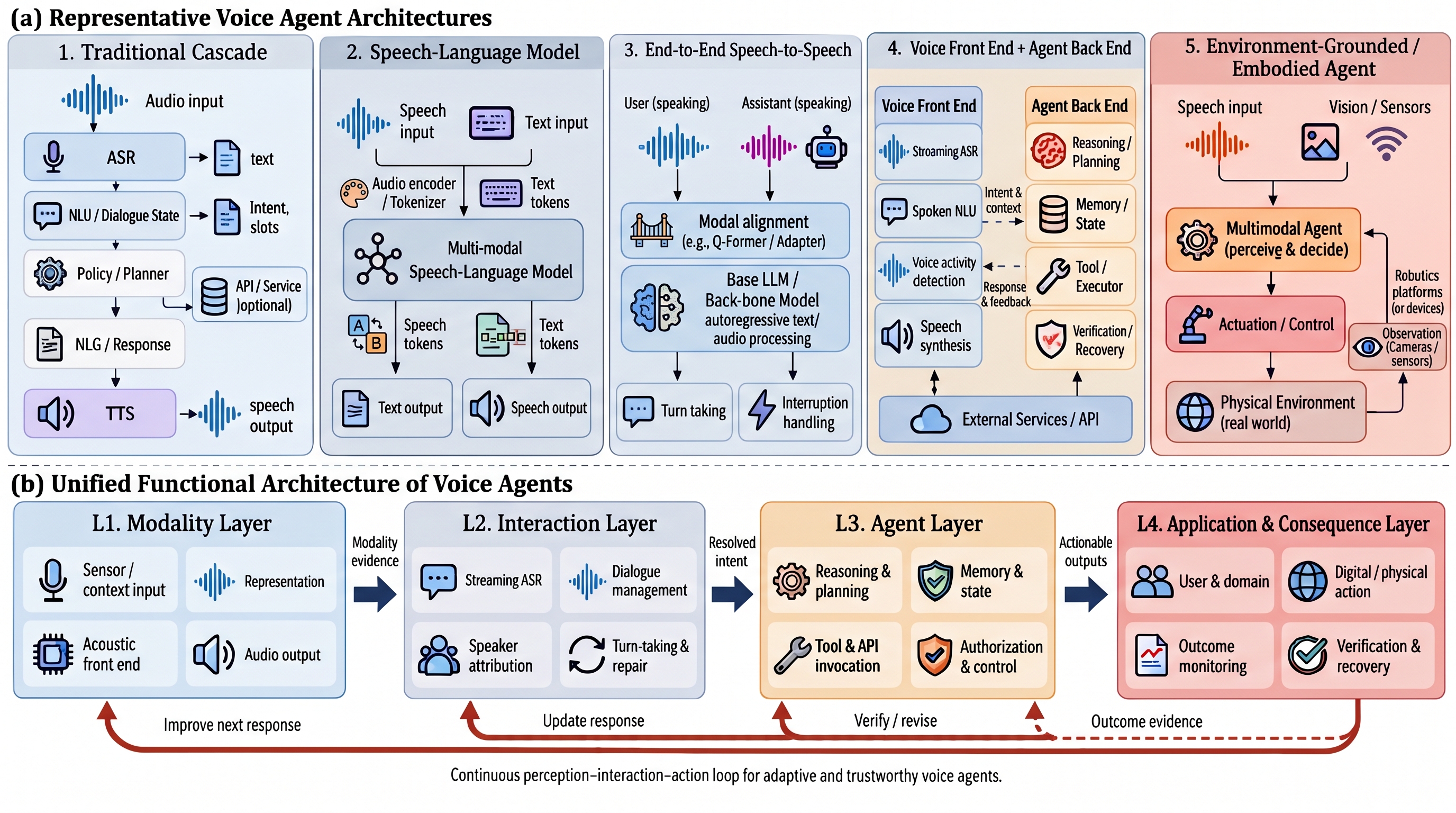}
\caption{Representative architectural paradigms and unified functional organization of voice agents. (a) Five representative architectural paradigms, ranging from traditional cascades and speech-language models to agent-integrated and environment-grounded systems. (b) A unified four-layer functional architecture spanning modality processing, interaction management, agent reasoning and control, and application-level consequences, with outcome-driven feedback supporting continuous adaptation and verification.}
\label{fig:evolution-stack}
\end{figure*}

The principal architectural differences lie in the interfaces through which information and control pass. Cascades commonly exchange text and structured dialogue state; speech-native systems can retain learned acoustic representations; duplex designs coordinate concurrent audio and control signals; and agent-oriented systems add calls, observations, and state updates. These are broad design families rather than mutually exclusive categories. Table~\ref{tab:architecture-families} compares their treatment of state and decision ownership, together with the interaction mechanisms that make those boundaries time-sensitive.

% Table II: consolidated architecture + real-time/full-duplex landscape.

\begin{table*}[tp]
\centering
\caption{Architecture and interaction design landscape. The upper block compares where state, decisions, and consequences are owned; the lower block summarizes mechanisms that make spoken interaction real-time and revisable. Reported numbers retain their original experimental scope.}
\label{tab:architecture-interaction}
\label{tab:architecture-families}
\label{tab:duplex}

\begingroup
\scriptsize
\setlength{\tabcolsep}{5pt}
\renewcommand{\arraystretch}{1.12}
\setlength{\extrarowheight}{1pt}
\setlength{\aboverulesep}{0pt}
\setlength{\belowrulesep}{0pt}
\arrayrulecolor{RoadRule}

\begin{tabularx}{\textwidth}{
  R{2.50cm} Z{0.95} Z{1.00} Z{1.05} R{3.00cm}
}
\toprule[0.8pt]
\rowcolor{RoadHead}
\textcolor{RoadInk}{\textbf{Pattern}} &
\textcolor{RoadInk}{\textbf{State / mechanism}} &
\textcolor{RoadInk}{\textbf{Decision or evidence boundary}} &
\textcolor{RoadInk}{\textbf{Systems implication}} &
\textcolor{RoadInk}{\textbf{Representative evidence}} \\
\midrule[0.5pt]

% ---------- A. Architectural envelopes ----------
\rowcolor{RoadHead!55!white}
\multicolumn{5}{l}{
  \textcolor{RoadInk}{\textbf{A. Architectural envelopes}}
} \\[3pt]
\midrule[0.3pt]

\textbf{Cascade}
&
text and explicit symbolic state exchanged between modules
&
dialogue manager and service layer; application observes outcome
&
visible interfaces, but accumulated latency and information loss
&
\cite{PAPER_E17F117D3387,PAPER_03CE3B0B842C}
\\[2pt]
\midrule[0.3pt]

\textbf{Speech--language model}
&
audio, text, or latent speech represented inside a shared model
&
shared model; tools and verification remain external when present
&
coherent cross-modal reasoning, with less visible ownership boundaries
&
\cite{PAPER_190E84406795,PAPER_C5043A3C2B57,PAPER_F09AF490D578}
\\[2pt]
\midrule[0.3pt]

\textbf{Native full duplex}
&
concurrent acoustic and semantic streams with generation-time control
&
joint generator; consequential action usually remains downstream
&
low-latency coordination, but interaction and action can become entangled
&
\cite{PAPER_1473A84B8A9D,PAPER_FAE46C323E2F,PAPER_65418243EE76}
\\[2pt]
\midrule[0.3pt]

\textbf{Layered voice agent}
&
streaming interaction front end plus planner and typed executor
&
planner proposes; executor commits; application or verifier observes
&
explicit commitment and consequence boundaries
&
\cite{PAPER_40A1EF47A5D0,PAPER_B421DE5BB9A8}
\\[2pt]
\midrule[0.3pt]

\textbf{Tool-augmented agent}
&
speech or text coupled to structured calls and observations
&
orchestrator plus external service
&
capability expands quickly; authorization and partial failure become first-class concerns
&
\cite{PAPER_766B59A1727C,PAPER_09240145263B,PAPER_9AD353416653}
\\[2pt]
\midrule[0.3pt]

\textbf{Embodied agent}
&
speech combined with vision, scene state, and control signals
&
multimodal policy or planner plus controller; sensors expose world state
&
consequence is observable, but reversibility and takeover are harder
&
\cite{PAPER_A40BBDA83DC5,PAPER_E9854767F763,EXTERNAL_ELLSA}
\\[3pt]
\midrule[0.5pt]

% ---------- B. Real-time and full-duplex interaction mechanisms ----------
\rowcolor{RoadHead!55!white}
\multicolumn{5}{l}{
  \textcolor{RoadInk}{
    \textbf{B. Real-time and full-duplex interaction mechanisms}
  }
} \\[3pt]
\midrule[0.3pt]

\textbf{Native duplex generation}
&
parallel codec streams or learned listen/speak control
&
concurrent listening and generation
&
native low-delay dialogue; interaction evidence does not by itself establish action closure
&
Moshi; SALM-Duplex; Freeze-Omni
\cite{PAPER_1473A84B8A9D,PAPER_FAE46C323E2F,PAPER_65418243EE76}
\\[2pt]
\midrule[0.3pt]

\textbf{Unified multimodal duplex}
&
autoregressive or shared multimodal voice modeling
&
concurrency inside a common model state
&
combines duplex dialogue with broader voice tasks
&
BayLing-Duplex; MinMo
\cite{PAPER_9DD954E00CAA,PAPER_F09AF490D578}
\\[2pt]
\midrule[0.3pt]

\textbf{Modular duplex control}
&
controller layered over an existing speech/language back end
&
turn policy remains separable from the back end
&
pluggable alternative to full-model retraining
&
FlexDuo; DuplexCascade
\cite{PAPER_30744A6D4C42,PAPER_9C69E07FE883}
\\[2pt]
\midrule[0.3pt]

\textbf{Synchronized speech + action}
&
speech, planning, turn labels, and structured calls share a clock
&
legal call timing and argument matching
&
exposes latency--action trade-offs; execution and authorization remain separate
&
DuplexSLA: 85.56\% at 0.648\,s vs. 91.33\% at 2.77\,s for a cascade
\cite{PAPER_09240145263B}
\\[2pt]
\midrule[0.3pt]

\textbf{Duplex benchmarks with tools}
&
overlap, disfluency, self-correction, and mocked APIs
&
interaction and bounded tool-call traces
&
connects timing to task evidence without claiming real side effects
&
Full-Duplex-Bench v2/v3: $F_1$ 0.876; pass@1 0.600
\cite{PAPER_B8CB31F1F27C,PAPER_9AD353416653}
\\[2pt]
\midrule[0.3pt]

\textbf{Interruption and recovery}
&
interruption scenarios plus recovery rubrics
&
post-interruption behavior is scored separately from ordinary task success
&
makes recovery an independent endpoint
&
FD-Bench; IHBench
\cite{PAPER_4F97D68B6C26,PAPER_2D9099D449E2}
\\[2pt]
\midrule[0.3pt]

\textbf{Endpointing / anticipation}
&
semantic completion, barge-in, multi-state decisions, and early prediction
&
timing of provisional versus committed interpretation
&
makes the commitment boundary explicit
&
\cite{PAPER_CB8E8D2C7474,PAPER_11D6F89D815C,PAPER_C60EBC02D547,PAPER_82CBE66BE086,PAPER_9DC02CA93758}
\\[3pt]

\bottomrule[0.8pt]
\end{tabularx}
\endgroup
\end{table*}

\textbf{Cascades and partially integrated pipelines.} A conventional cascade separates the acoustic front end, recognition, language understanding, dialogue management, back-end services, and synthesis. Intermediate outputs make errors easier to locate, provided the stages are observable. The costs are accumulated latency, error propagation, and the loss of prosody or overlap when speech is reduced to text. Partial integration can address some of these costs without removing every boundary. Conversational synthesis conditions generation on dialogue history~\cite{PAPER_79AAF689FC20}; streaming understanding extracts intent and slots from partial input~\cite{PAPER_C69F6F77FA7A}; duplex wrappers add concurrent listening and speaking to sequential back ends~\cite{PAPER_30744A6D4C42}; and hybrid controllers separate rapid interaction handling from slower reasoning~\cite{PAPER_13438B5C64C7}.

\textbf{Token-based and speech-language architectures.} Many speech-native designs replace a text-only interface with learned discrete units, codec tokens, or direct audio representations. Discrete-unit and two-pass approaches use learned speech units for generation~\cite{PAPER_BE36882FCC21,PAPER_A729F1BBE19F,PAPER_B591F4760692}. AudioPaLM consumes and generates speech within a shared model~\cite{PAPER_190E84406795}; SALMONN-omni combines streaming encoding and decoding without a codec bottleneck~\cite{PAPER_C5043A3C2B57}; and unified multimodal models extend speech processing across interaction tasks~\cite{PAPER_F09AF490D578}. Retaining acoustic information can preserve cues that a plain transcript would discard and may reduce interface overhead. The trade-off is less visible internal decision-making, particularly when planning or tool use is folded into the same model. An intermediate text stream, as in Moshi, offers some additional visibility without making the entire action path explicit~\cite{PAPER_1473A84B8A9D}.

\textbf{End-to-end and duplex architectures.} Duplex systems treat listening and speaking as overlapping activities rather than strictly alternating turns. Two-channel generative dialogue models established one formulation~\cite{PAPER_1305D1F3BECF}; Moshi models parallel speech streams~\cite{PAPER_1473A84B8A9D}; SALM-Duplex learns listen-or-speak behavior~\cite{PAPER_FAE46C323E2F}; and Freeze-Omni learns a duplex speech interface while retaining a frozen language model~\cite{PAPER_65418243EE76}. Turn-level interleaving provides a different balance between integrated modeling and interaction control~\cite{PAPER_760547F36D09}. In these architectures, turn-taking can become part of generation rather than a separate decision made before response production. That improves the prospects for fluid dialogue, but it can also make a fluent yet incorrect response harder to trace to a particular intermediate decision.

\textbf{Voice front ends with agent back ends.} A common route to task-oriented voice interaction is to pair a front end responsible for endpointing, concurrency, and turn control with a back end responsible for reasoning and execution. AnovaX describes a speech front end, optional planner, typed executors, safety layer, and lifecycle event bus~\cite{PAPER_40A1EF47A5D0}; AURA similarly separates understanding, reasoning, and automated tool use~\cite{PAPER_B421DE5BB9A8}. Domain-oriented systems adopt related arrangements for form filling, search, and support~\cite{PAPER_3FD26CDF1D8B,PAPER_838318340492,PAPER_0BB1928924AB,PAPER_E4FD61E964C3}. Explicit action interfaces help localize failures and constrain what can be attempted. Those constraints can also shape reported reliability: a narrow executor has fewer opportunities to fail. More fundamentally, a well-formed tool call may still be unauthorized, rejected by a service, or executed without producing the intended state.

\textbf{Grounded and embodied architectures.} Physical environments make the interaction between speech and action particularly visible. DOROTHIE interleaves dialogue moves with navigation decisions~\cite{PAPER_A40BBDA83DC5}; other robotic systems use omni-modal context for proactive manipulation~\cite{PAPER_E9854767F763} or support low-latency multimodal interaction on a service robot~\cite{PAPER_4C2C38D58510}. Systems that divide attention between speech and action experts also expose a practical cost of concurrency: speaking while acting can degrade long-horizon task performance~\cite{EXTERNAL_ELLSA}. Grounding requires references to be resolved against the scene, repairs to arrive at the right time, and control to remain available for human intervention. Several representative evaluations still rely on simulation and scripted or generated interactions, limiting what they establish about real-world operation~\cite{PAPER_A40BBDA83DC5,PAPER_A8347D87D2F8}.

\subsection{Distribution of Intelligence and Control Across the Stack}
\label{sec:evolution-distribution}

The architectural families differ in how they assign listening, deciding, remembering, and acting. Figure~\ref{fig:evolution-stack}(b) organizes the recurring functions into modality, interaction, agent, and application-and-consequence layers. Table~\ref{tab:architecture-families} shows how a given function may reside in a single model or be divided among an interaction front end, reasoner, controller, executor, and memory store. This placement affects both system behavior and the evidence available to evaluate it.

Three broad arrangements recur. In \emph{centralized} designs, one model handles audio input and output, leaving many perceptual and conversational decisions internal~\cite{PAPER_1473A84B8A9D,PAPER_F09AF490D578}. Fewer explicit handoffs can simplify interaction, but an observer may be unable to distinguish a missing decision from one that was made incorrectly. \emph{Layered} systems pair a fast interaction front end with a slower reasoning or action back end~\cite{PAPER_30744A6D4C42,PAPER_13438B5C64C7}. This keeps action interfaces accessible and helps allocate latency, provided the interface carries corrections and goals faithfully. In \emph{distributed} systems, perception, interaction control, reasoning, memory, and execution have more distinct owners~\cite{PAPER_40A1EF47A5D0,PAPER_B421DE5BB9A8}. Components can be replaced and failures more readily localized, but each additional contract can lose information about identity, uncertainty, or authorization.

These choices increasingly extend to the device itself. Qualcomm's 2026 Snapdragon 8 Elite Gen 6 platforms describe an enhanced Sensing Hub, on-device context processing, and AI Engine resources for agentic workloads. The Extreme platform also advertises AI Voice Bubble for voice--ambient separation and support for on-device 30B+ mixture-of-experts models~\cite{QUALCOMM_GEN6_2026,QUALCOMM_EXTREME_GEN6_2026}. Secondary reporting quotes Qualcomm on small-model execution in the Sensing Hub and a complete voice-in/voice-out agent on the platform~\cite{TECHCRUNCH_GEN6_2026}. These remain vendor-reported capabilities, not independent demonstrations of auditory agency. They do, however, move questions of latency, privacy, personalization, and control ownership across the device--cloud boundary.

Centralized systems may reduce interface overhead, while layered and distributed systems offer more places to observe authorization, execution, and verification. This distinction matters when the component that makes a decision also reports its success: the report is not an independent observation of the outcome. Grounded evaluation and explicit execution interfaces illustrate why observable boundaries matter~\cite{PAPER_535B4C25EDC0,PAPER_40A1EF47A5D0}. The useful design question is which boundaries need to remain inspectable for a particular application, rather than whether the system is labeled integrated or modular.

\subsection{Architectural Trade-offs}
\label{sec:evolution-tradeoffs}

No architectural family is preferable in every setting. The relevant trade-offs depend on the latency a task can tolerate, the acoustic information it needs, and how much control and evidence must remain accessible when an action has consequences.

\textbf{Latency.} Cascades incur delays across stages, although some processing can run in parallel; integrated models remove handoffs but still pay for generation. Reported latency is meaningful only when its endpoints are clear. Model-chunk latency, first-audio latency, endpoint-to-response latency, and microphone-to-action latency measure different intervals and should not be compared directly~\cite{PAPER_09240145263B,PAPER_9AD353416653}.

\textbf{Modularity and inspectability.} Modular designs can expose state and policy as separate, inspectable objects~\cite{PAPER_E17F117D3387}. Centralized models offer fewer such boundaries unless they are deliberately instrumented. Shared research toolkits such as ESPnet support reproducible component-level speech experiments~\cite{watanabe2018espnet}, but tracing an agent's consequential decisions requires visibility beyond the speech model itself.

\textbf{Acoustic information preservation.} Text interfaces discard prosody, overlap, and other non-lexical cues unless these are encoded separately. Direct speech-to-speech systems can retain learned speech representations across the interface~\cite{PAPER_BE36882FCC21}; codec-free speech-language models keep audio in the input and output path through a different design~\cite{PAPER_C5043A3C2B57}. Neither choice guarantees that the information will be used correctly downstream.

\textbf{Reasoning capability and memory integration.} Placing dialogue context close to a reasoning model can simplify the use of incomplete or evolving information~\cite{PAPER_CB6C6E4E2062}. Persistent memory raises different requirements: stored preferences and histories may need to be inspected, corrected, or deleted independently of the model that uses them~\cite{PAPER_FFAF306C931D,PAPER_07DE94B0DB13}. A convenient location for inference is not necessarily the right location for long-lived user state.

\textbf{Tool integration and deployment cost.} External services extend what a voice agent can do, but they add dependencies that must be managed alongside model computation. Speech-to-function and synchronized-action systems expose the interfaces through which structured calls are produced~\cite{PAPER_766B59A1727C,PAPER_09240145263B}. In deployed systems, latency, availability, permissions, and service configuration can be as consequential as the accuracy of the call itself.

\textbf{Failure isolation, safety, and controllability.} Distributed architectures can make faults easier to isolate and permit unsafe components to be disabled or replaced~\cite{PAPER_BA00BDF2544F}. Greater integration places more responsibility within the model and can complicate controlled degradation. For consequential actions, an explicit authorization boundary remains valuable even when perception, reasoning, and response generation share a common model.

\section{Interaction Intelligence: From Streaming Speech to Full-Duplex Dialogue}
\label{sec:interaction}

The architectural choices discussed in the previous section become especially consequential during live conversation. A voice agent must interpret speech as it arrives, determine who is speaking and whether the utterance is meant for it, and respond to interruptions or corrections that may arrive after processing has begun. We use \emph{interaction intelligence} to describe the perceptual and conversational capabilities that support this process: perception, attribution, incremental understanding, turn prediction, coordination, and repair. These functions need not operate in a fixed sequence. Their outputs evolve over time, and a revision at one stage may affect reasoning, memory, or an action already being prepared elsewhere in the system.

\begin{figure*}[t]
    \centering
    \includegraphics[width=\textwidth]{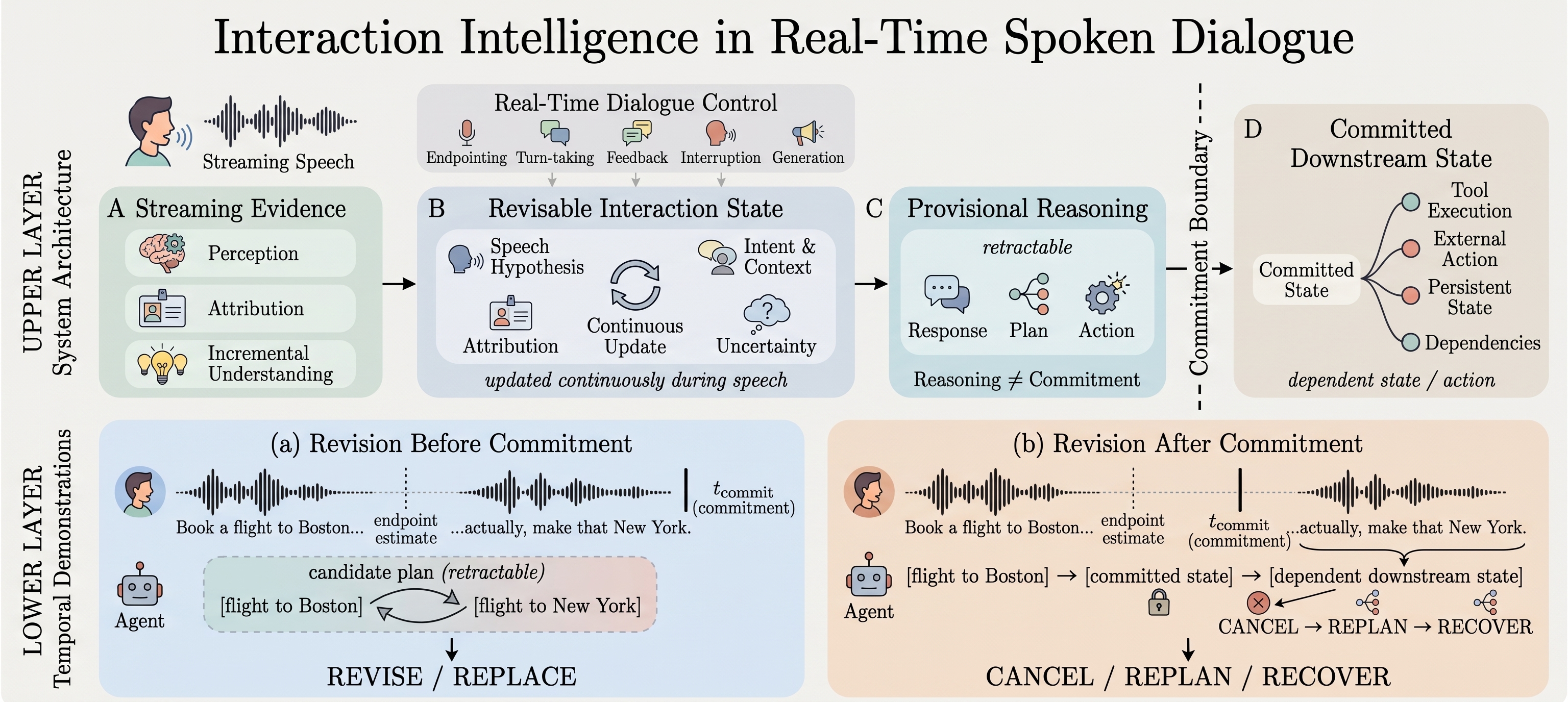}
    \caption{Overview of interaction intelligence in real-time spoken dialogue. Streaming speech evidence updates a revisable interaction state through perception, attribution, and incremental understanding. Real-time dialogue control operates on that state while responses and plans may be prepared provisionally before commitment. Later spoken corrections can trigger repair and replanning, whereas crossing the commitment boundary makes downstream state consequential. Stopping speech output and cancelling a prepared action are therefore distinct control decisions.}
    \label{fig:interaction-overview}
\end{figure*}

Figure~\ref{fig:interaction-overview} provides an overview of this organization. Streaming speech evidence is continuously incorporated into a revisable interaction state through perception, attribution, and incremental understanding. Real-time dialogue-control functions---including endpointing, turn-taking, feedback, interruption handling, and generation---operate on that evolving state, while candidate responses or plans may be prepared provisionally before commitment. Later spoken evidence can still trigger repair and replanning. The commitment boundary marks the point after which downstream components may begin to depend on the current interpretation, making interaction-level interruption and action-level cancellation distinct control decisions.

\subsection{Perceiving: Front Ends and Recognition}
\label{sec:interaction-perceive}

Under close-talk, single-speaker conditions, the acoustic front end may be almost invisible to the rest of the system. Its role becomes much more apparent when the agent's own playback, room reverberation, or competing speakers enter the microphone signal. Echo cancellation and personalized enhancement attenuate playback and interfering acoustic components \cite{PAPER_92B2C1256F1A,PAPER_A5EAA3B68864}, while voice activity detection establishes whether speech is present. Speaker-specific activity detection further helps distinguish an intended interruption from residual playback or background conversation \cite{PAPER_D9E5B1683133}. Errors at this stage can change the course of an interaction before recognition has even begun.

Recognition remains another important source of upstream uncertainty. Conformer introduced a convolution-augmented Transformer architecture for speech recognition~\cite{gulati2020conformer}, while Whisper demonstrated the value of large-scale weak supervision~\cite{radford2023robust}. Work on accents, dialects, code-switching, and low-resource speech includes joint diarization--recognition pipelines, multilingual evaluation, accent-robust detection, and systems developed for particular language varieties \cite{PAPER_3FE31C58A49B,PAPER_640D8E073990,PAPER_5C4380166688,PAPER_A8FB29F113AF,PAPER_F2778FA85677}. Federated and on-device approaches also seek to improve recognition without requiring all audio to be centralized \cite{PAPER_8902318ADCC4,PAPER_E5C507BB2EE0}. Reported error rates, however, depend heavily on the domain, accent distribution, and noise conditions of the test set. More importantly for an agent, recognition errors have unequal consequences: a mistaken proper noun, number, or slot value may alter the intended operation even when the rest of the utterance is transcribed correctly. This sensitivity helps explain the continued interest in entity-aware and retrieval-augmented methods \cite{PAPER_76F78E88F459,PAPER_7C61510F1269}.

\subsection{Attributing: Who Spoke, and to Whom}
\label{sec:interaction-attribute}

Recognizing the words is only part of interpreting a spoken request. In a shared acoustic environment, the system must also establish whether the utterance was directed at it and which speaker produced it. Device-directedness addresses the former question, helping prevent conversations among people from being mistaken for commands or interruptions \cite{PAPER_40CE2E5F8957}. Diarization and multi-speaker separation provide evidence about speaker identity and temporal ownership of the signal; efficient on-device diarization shows that such processing can be performed under local resource constraints \cite{PAPER_E5C507BB2EE0}. Addressee recognition in multi-party dialogue and benchmarks for interactional privacy make these distinctions explicit evaluation targets \cite{PAPER_B49F83A8B5BB,PAPER_49C7AB780EBD}.

Tools such as pyannote.audio support the temporal assignment of speech segments to speakers~\cite{bredin2020pyannote}, and x-vector embeddings provide representations for speaker recognition~\cite{snyder2018x}. Their significance changes when the system can take actions. Misattribution in a question-answering exchange may produce an irrelevant reply; in an agent, it may cause a request to be carried out on behalf of the wrong person. Even a correctly recognized speaker may lack authority to initiate a particular operation. The literature therefore distinguishes speaker identity, verification, and identification \cite{PAPER_E648C7BE210D}; none of these, on its own, establishes authorization. We return to this boundary in Section~\ref{sec:human}.

\subsection{Understanding and Predicting: Endpoints and Commitment}
\label{sec:interaction-understand}

Spoken language understanding has moved beyond extracting intent and slots from a single best transcript. Confusion networks retain alternative recognition hypotheses for dialogue state tracking \cite{PAPER_03CE3B0B842C}; joint intent-and-slot models assign labels to streaming input \cite{PAPER_C69F6F77FA7A}; and temporally expressive networks represent changes in belief state across turns \cite{PAPER_86E0C27FBC57}. Intent accuracy, slot $F_1$, and joint goal accuracy remain useful measures in bounded domains, but they do not establish that a request was fulfilled appropriately. The intended operation may be understood correctly yet remain unauthorized, while an apparently correct slot value may still be used in the wrong action.

A related challenge is deciding when enough of the utterance has arrived to proceed. Pause-based endpointing forces a compromise between interrupting a slow speaker and delaying a response. Turn-completion projection uses incrementally available conversational evidence to predict when the speaker is likely to finish~\cite{ekstedt2021projection}. Semantic endpointing incorporates partial meaning into that decision \cite{PAPER_CB8E8D2C7474}, and streaming endpointers can be trained with delayed labels to support decisions before the acoustic boundary is fully clear \cite{PAPER_9DC02CA93758}. More aggressive anticipation predicts turn endings before they occur: in the reported Unmute integration, predictions up to 2.56\,s ahead reduced average response latency by approximately 505\,ms, with 28.4\% speculative-computation overhead \cite{PAPER_82CBE66BE086}.

An endpoint estimate is nevertheless different from a decision to commit. We denote the estimated acoustic or semantic endpoint by $t_{\mathrm{end}}$ and the commitment time by $t_{\mathrm{commit}}$. The former concerns whether the speaker has finished; the latter identifies when the system allows its current interpretation to govern persistent state or consequential downstream behavior. A system may reason speculatively before $t_{\mathrm{end}}$, provided its intermediate state remains retractable. Alternatively, it may wait beyond the endpoint while resolving speaker attribution, ambiguity, or authorization. The important design choice is how uncertainty is carried between early interpretation and commitment.

This distinction matters whenever the speaker changes course. A late correction may invalidate an argument in a pending tool call, alter a plan, or require an earlier state update to be reconsidered. The interaction layer must be able to pass that revision downstream rather than treating the latest transcript as an isolated replacement. Here gives a concrete timing example of the distinction summarized in Figure~\ref{fig:interaction-overview}. Before $t_{\mathrm{commit}}$, the candidate interpretation and plan remain retractable and can be revised directly; afterward, other components may already depend on them, making cancellation, replanning, or recovery necessary.

Turn-taking has long been central to spoken dialogue research, with control strategies progressing from heuristics toward learned policies~\cite{raux2012optimizing,skantze2021turn,khouzaimi2016reinforcement}. The timing of human conversational turns provides an important constraint on these systems~\cite{levinson2016turn}. Prosodic and lexical cues help predict when a response is expected \cite{PAPER_DE1FE99D0F91} and when an agent may begin speaking \cite{PAPER_0BE6D6719412}. Later models combine acoustic, lexical, and contextual information: one multimodal approach reports 0.826 macro-$F_1$ for barge-in on a real interactive-voice-response corpus \cite{PAPER_C60EBC02D547}, while a four-state classifier distinguishes pause, backchannel, interruption, and continuation with approximately 95.75\% mean accuracy on its reported test set \cite{PAPER_11D6F89D815C}. Backchannel prediction has likewise progressed from voice-activity projections \cite{PAPER_016FC4509054} to acoustic--language-model fusion \cite{PAPER_D49E9CF62CAF} and cross-lingual transfer studies \cite{PAPER_891085526945}. The distinction between acknowledgement and interruption is particularly important: a system that stops speaking whenever it hears a brief response may detect speech correctly while managing the conversation poorly.

Barge-in brings turn coordination back to the question of attribution. When someone speaks over the agent, the controller must decide whether the speech is addressed to the system, whether its output should stop, and whether the new utterance changes the current goal. Device-directedness helps address the first question \cite{PAPER_40CE2E5F8957}, while personalized activity detection and acoustic processing reduce false triggers from playback and interference \cite{PAPER_D9E5B1683133,PAPER_A5EAA3B68864,PAPER_92B2C1256F1A}. Stopping the audio output is only one possible response. If the interrupted utterance was linked to ongoing reasoning or a prepared operation, the controller must also determine whether that work can continue, be revised, or be cancelled.

Barge-in is one form of a broader overlap problem. Speakers may talk simultaneously, listeners may acknowledge without taking the floor, and corrections may begin before a preceding utterance has ended~\cite{heldner2010pauses,khouzaimi2015turn}. Frame-synchronous models, streaming context, and context-aware benchmarks now treat turn-taking as a joint acoustic and linguistic task \cite{PAPER_0E189C354089,PAPER_1B4896BFDF27,PAPER_451C2CE21206,PAPER_6953C626B3AE}. Other studies examine the meaning of interruptions, not just their timing \cite{PAPER_A38532826B7B,PAPER_4C27C22ED486,PAPER_5A7AEFF2BD2C}, while interference-resilient fusion and user-role control address robustness under more complex conditions \cite{PAPER_7CDC981954ED,PAPER_546051179C13}. Together, these results show why conversational coordination cannot be reduced to detecting whether a microphone channel is active.

\subsection{Generating Speech Inside the Loop}
\label{sec:interaction-generate}

Speech generation brings another set of objectives into the interaction loop. Recognition is commonly evaluated through transcription errors, whereas synthesis must account for intelligibility, naturalness, speaker identity, and prosodic fidelity. Tacotron introduced end-to-end prediction of speech representations from text~\cite{wang2017tacotron}; Tacotron~2 combined mel-spectrogram prediction with a WaveNet vocoder~\cite{shen2018natural}; and HiFi-GAN made high-fidelity waveform generation more efficient~\cite{kong2020hifi}. Conversational synthesis subsequently incorporated dialogue history into acoustic generation \cite{PAPER_79AAF689FC20}, while reference-based measures of prosody and rhythm provided alternatives to relying solely on mean opinion scores \cite{PAPER_777467AF8540}. These advances improve the voice heard by the user, but they do not by themselves determine how the system should respond to a changing conversation.

Voice also carries affect and speaker identity. StyleTTS~2 combines style diffusion with adversarial training and large speech language models to improve naturalness and speaker adaptation~\cite{li2023styletts}. Emotion transfer has been explored in synthesis and zero-shot voice conversion using speech-language-model representations \cite{PAPER_717DDA83DB83,PAPER_FCBABE30F7D4}; longer-horizon expressivity control has been studied in embodied settings \cite{PAPER_1948D2F1F654}; and affect-annotated conversational corpora support work on empathetic dialogue \cite{PAPER_C2C0128ACA13}. Other approaches integrate prosody with multi-agent reasoning, encode affective structure before response generation, or train empathetic spoken chatbots around frozen language models \cite{PAPER_E3C23C0C4A4E,PAPER_4A8AAC6B4409,PAPER_170D457B6045}. Such work makes generated speech more expressive, although an appropriate vocal style is not sufficient evidence of emotion understanding or safe behavior in a high-stakes interaction.

Verification has also begun to appear within the speech-generation process. CycleSpeech links instruction-controlled synthesis with paralinguistic understanding through a structured voice profile. In its forward cycle, a profile inferred from synthesized speech is compared with the intended vocal attributes; in its backward cycle, a profile inferred from real speech is used to reconstruct the original speaking style. Trained on 20,046 bilingual examples, the method reports higher instruction-match accuracy than its Step-Audio-2-mini baseline in both Chinese and English \cite{CYCLESPEECH_2026}. The verification here concerns the speech signal and its intended attributes. It does not establish that an external action was executed or that its outcome was correct.

Speech translation provides a related perspective because its evaluation has developed around several competing objectives. Cascaded systems retain intermediate representations that can be inspected stage by stage, while direct approaches generate target speech without an explicit transcript \cite{PAPER_A224F6D9B6D8,PAPER_BE36882FCC21,PAPER_A729F1BBE19F,PAPER_B591F4760692,PAPER_79C802D98E40}. BLEU and ASR-BLEU measure aspects of content fidelity; mean opinion scores and speaker similarity assess speech quality; and real-time factor or lag captures timing. Simultaneous translation makes the trade-off especially visible: beginning the target speech before the source utterance ends can reduce delay but may compromise adequacy. New benchmarks therefore examine expressiveness, long-form simultaneous translation, and multi-turn quality alongside conventional content measures \cite{PAPER_5BBBCAA1CFCE,PAPER_CDFE5E5493EF,PAPER_447A023EBD30,PAPER_9B5A0DFE07A6}. Multilingual studies further test performance across accents, low-resource languages, vocalization patterns, and typological conditions \cite{PAPER_A8FB29F113AF,PAPER_4B62BAAC24BC,PAPER_3B58C15C3C82}. Strong aggregate multilingual performance can conceal substantial differences among languages, which become operationally important when an agent relies on recognition or generation to complete a task.

Speech generation is consequently better measured at the signal level than at the level of conversational consequences. Naturalness, fidelity, translation quality, and latency have established evaluation procedures. Less well understood is how generated speech affects coordination and action: whether output stops appropriately during an interruption, whether vocal style changes a user's interpretation of a request, or whether a confident-sounding response masks an unsuccessful operation. These questions become more pressing when listening and speaking occur concurrently.

\subsection{Full-Duplex Technology}
\label{sec:interaction-duplex}

Full-duplex dialogue allows a system to receive user audio while its own speech is active and to decide whether to continue, stop, yield, backchannel, or resume. It involves ongoing interaction control, beyond simultaneous translation, low generation latency, or the mere production of two audio streams. A model may exhibit any of those properties without reliably managing the conversational floor.

\textbf{Modeling concurrency.} The two-channel formulation predates recent speech foundation models. Generative spoken dialogue modeling showed that both sides of a conversation could be represented as parallel streams, including spontaneous overlap \cite{PAPER_1305D1F3BECF}. Moshi develops this approach with parallel semantic and acoustic codec streams and an internal text stream, reporting approximately 160\,ms theoretical and 200\,ms practical latency \cite{PAPER_1473A84B8A9D}. The text stream, described as an inner monologue, offers an intermediate representation alongside continuous acoustic generation. Other systems make the listen-or-speak decision an explicit learning objective: SALM-Duplex learns when to listen and speak \cite{PAPER_FAE46C323E2F}, while Freeze-Omni learns a duplex speech interface around a frozen language model \cite{PAPER_65418243EE76}. GLM-4-Voice and longer-context models pursue end-to-end conversational speech \cite{PAPER_42139A057A67,PAPER_30C552706E67}; SALMONN-omni connects streaming encoding and synthesis without a codec bottleneck \cite{PAPER_C5043A3C2B57}; and modular alternatives add duplex control to an otherwise conventional back end \cite{PAPER_30744A6D4C42}. In integrated designs, the choice to continue or yield becomes part of generation itself. This can align timing decisions with the representation used to produce speech, although it may leave less explicit evidence of how those decisions were made.

\textbf{Benchmarks and what they changed.} Evaluation has followed the growing complexity of these interactions. Early turn-taking metrics on controlled audio \cite{PAPER_A4706CE702D0} were extended to overlap handling \cite{PAPER_760BC4ACD4D7}, automated multi-turn examination \cite{PAPER_B8CB31F1F27C}, and disfluent human recordings paired with mock tool APIs \cite{PAPER_9AD353416653}. Complementary benchmarks generate interruption-rich conversations under controlled timing and noise \cite{PAPER_4F97D68B6C26}, isolate recovery after interruption \cite{PAPER_2D9099D449E2}, examine human-likeness \cite{PAPER_FF1108C3BF58,PAPER_1B06772DA179}, or evaluate naturalness in streaming speech-to-speech systems \cite{PAPER_EBFD723A1FA4}. Talking Turns offers another instrument for assessing turn-taking behavior in audio foundation models~\cite{arora2025talking}.

These benchmarks increasingly report separate measures for taking the turn, interruption, stopping and resuming, recovery, and response quality. Full-Duplex-Bench-v3 illustrates the need for this separation. Across 100 recordings from 12 speakers and four mock API domains, one real-time model achieves tool-selection $F_1$ of 0.876 and pass@1 of 0.600, with a 96\% take-turn rate and 6.89\,s task latency. A faster competitor responds in 4.25\,s but takes the turn in only 78\% of cases, whereas a cascaded system takes every turn at 10.12\,s and has a higher interruption rate \cite{PAPER_9AD353416653}. A single latency or success score would obscure these differences.

\textbf{Duplex control with structured action.} The connection to agency becomes more direct when structured actions share the speech timeline. DuplexSLA synchronizes user audio, assistant audio, and action output on a common 160\,ms clock. It produces planning text, JSON calls, and interruption or backchannel labels while speech continues. On 900 deterministic tool cases, it reports 85.56\% mean accuracy at 0.648\,s delay, compared with 91.33\% at 2.77\,s for a cascade; turn-decision accuracy ranges from 93.33\% to 99.33\% across states \cite{PAPER_09240145263B}. This is evidence of coordinated speech and structured-call generation under the benchmark's timing constraints. Its deterministic function-and-argument matching criterion does not establish authorization, real service execution, or verification of an external state change.

A recent embodied system makes the timing of commitment more tangible. MIRA combines streaming speech, turn arbitration, and co-speech motion on an Astribot S1 robot. Its CORTEX controller separates rapid interruption handling from deliberative turn decisions, while ROSCO generates motion using Receding-Horizon Prefix Commitment: it predicts farther ahead than it releases, allowing only a short, cancellable motion prefix to enter physical execution \cite{MIRA_2026}. Once motion has been dispatched, it cannot be revised as freely as an internal text plan. Limiting the committed prefix therefore provides a concrete form of temporally revisable control. The reported experiments cover real-robot streaming and interruption handling, but do not establish the full chain from authorization through external-state verification to compensation. Table~\ref{tab:duplex} compares representative interaction systems and benchmarks; MIRA is discussed separately because it appeared after the frozen search cutoff.

\subsection{Repair: From Stopping to Reverification}
\label{sec:interaction-repair}

A system may detect an interruption promptly yet recover poorly from its consequences. In an evaluation spanning 45 synthetic state-machine conversations, 10 enterprise domains, 27 model configurations, and six interruption types, the highest task-fulfillment rate and the highest recovery-quality pass rate came from different systems. Agreement between automated judges and humans was also modest despite high agreement among the judges themselves \cite{PAPER_2D9099D449E2}. Other studies find that repeated requests are often recognized as recovery situations without being repaired consistently \cite{PAPER_4F97D68B6C26}, or that a model commits to an obsolete intent before a self-correction arrives \cite{PAPER_9AD353416653}. Recovering the dialogue, stopping the speech output, and cancelling a downstream action are distinct outcomes.

Work on disfluency and speech repair provides a linguistic foundation for understanding such corrections~\cite{shriberg1994preliminaries,heeman1999speech}. For a voice agent, however, repair must reach the state used to carry out the task. A mid-utterance correction may change the goal, the arguments of a pending call, or the plan that the system has already prepared. Clarification in dialogue policies and recovery in agent frameworks address parts of this problem \cite{PAPER_EC0445D89B24,PAPER_913915B00C96}, but the cited evaluations examine them separately. Current interruption benchmarks therefore provide clearer evidence about whether a system stopped or resumed than about whether it correctly revised the intended operation.

\subsection{What Interaction Intelligence Does Not Settle}
\label{sec:interaction-limit}

The interaction layer has a comparatively mature set of measurements. Recognition, speaker attribution, endpointing, state tracking, and turn control can be evaluated with dedicated metrics and known failure cases. Their outputs also affect decisions beyond the interaction layer. Attribution matters when speakers have different permissions; endpointing influences when a provisional interpretation becomes available; and barge-in can require ongoing computation or a prepared action to be revised. The quality of these upstream decisions is important, but it does not by itself determine the legitimacy or correctness of the resulting action.

Full-duplex technology addresses how people and systems share the conversational floor while both can speak. A voice agent must additionally decide when to rely on an interpretation, whether a requested operation is authorized, and what to do after an action has changed external state. Those responsibilities call for explicit mechanisms for state, reasoning, memory, execution, observation, and recovery. The next section examines that agent loop and the interfaces through which it receives provisional evidence from spoken interaction.

\section{Agentic Intelligence: From Spoken Intent to Consequential Action}
\label{sec:agentic}

Spoken interaction becomes consequential when an interpretation is retained, used to prepare an operation, or passed to a system that can alter persistent or external state. Reasoning and tool calling provide important capabilities, but neither alone establishes that the user's goal has been carried through planning, execution, observation, and recovery~\cite{mialon2023augmented,xi2025rise}. We use \emph{agentic intelligence} to describe this goal-directed dimension of a voice agent. It cannot be separated entirely from interaction intelligence: a plan may depend on speech that is still being revised, and the authority to act depends on who made the request and under what circumstances. The problem is to preserve these dependencies as an interpretation becomes a decision and, eventually, an observable consequence. Table~\ref{tab:reasoning-memory-tools} summarizes the interfaces involved, together with the human-control obligations they create.

\subsection{State and Spoken Reasoning}
\label{sec:agentic-state}

Reasoning occupies different positions in current voice-agent architectures. It may take the form of a policy over explicit dialogue state, a language model behind a speech front end, or a speech-native model that processes acoustic representations directly. Other designs use a separate planner or delegate a request to an external agent. These arrangements differ not only in flexibility and latency, but also in what can be inspected when a decision goes wrong.

Classical dialogue systems made the decision process relatively explicit. Plato separates dialogue state, policy, and services into inspectable components \cite{PAPER_E17F117D3387}; reinforcement-learned policies operate over tracked belief states \cite{PAPER_A2DB3D2CB7CB,PAPER_EC2365EC8FAE}; and interaction-quality estimation has been investigated as a policy-learning signal rather than only a retrospective rating \cite{PAPER_EF55C422938D}. With an explicit state representation, an incorrect belief update can be distinguished from a poor policy choice or an unsuccessful service operation. The trade-off is that fixed representations may struggle to accommodate interrupted requests, reversals of earlier constraints, or references that depend on a longer conversation.

Connecting a speech front end to a language-model reasoner offers greater flexibility. Audio models can be coordinated by a language model \cite{PAPER_E3E2EEA2C7C7}, intermediate reasoning can be encouraged through chain-of-thought training \cite{PAPER_CB6C6E4E2062}, and streaming systems can initiate reasoning from semantic events before an utterance is complete \cite{PAPER_13438B5C64C7}. Yet the arrangement also carries uncertainty forward. Where the model relies on a transcript, a recognition error may change the inferred goal; where it produces a plausible plan, that plan still requires evidence of correctness. Deliberation adds latency as well, so the choice of reasoning strategy affects the interaction budget discussed in Section~\ref{sec:interaction} \cite{PAPER_8B3C1EFAB4CB}.

Speech-native models avoid making a transcript the mandatory interface between perception and reasoning \cite{PAPER_C5043A3C2B57,PAPER_F09AF490D578}. Some duplex architectures also retain an internal text stream while acoustic generation continues \cite{PAPER_1473A84B8A9D}. These designs can make use of prosody, background sounds, and overlapping speech that a transcript may not preserve. Their reported evaluations, however, address different questions. Spoken reasoning benchmarks examine inference, colloquial language, and paralinguistic understanding \cite{PAPER_CA77D0B34721}, whereas tool-oriented tests include monitoring and trigger detection \cite{PAPER_913915B00C96}. Neither kind of result, on its own, establishes calibrated confidence or appropriate abstention under uncertain acoustic evidence.

The architectural choice thus determines what decision evidence survives. Explicit policies expose intermediate state; integrated models can make the path from perception to response harder to inspect; and separate planners recover some visibility through structured outputs, at the cost of additional interfaces. Whichever design is used, consequential decisions must remain traceable to the information on which they were based.

\subsection{Planning and Replanning}
\label{sec:agentic-planning}

Planning turns an interpreted request into a course of action. Task-oriented frameworks decompose goals into structured operations for an executor \cite{PAPER_40A1EF47A5D0,PAPER_B421DE5BB9A8}. Situated agents must also revise their behavior as the environment changes. DOROTHIE, for example, combines dialogue decisions and navigation in a driving simulator, where what the agent says and what it does must respond to the same evolving situation \cite{PAPER_A40BBDA83DC5}.

Speech makes a plan unusually vulnerable to revision. A speaker may clarify a reference, change a constraint, or interrupt while an operation is being prepared. Dialogue-policy work studies when to seek clarification \cite{PAPER_EC0445D89B24}, but clarification is useful only if its result reaches the dependent plan. The studies considered here examine clarification and recovery in different settings; they provide more limited direct evidence of the complete sequence from spoken correction to replanning and cancellation of an obsolete operation.

That sequence becomes harder to manage once a proposal crosses into execution. An internal plan can often be revised without external effects; an operation already dispatched may instead require cancellation, a compensating action, or human intervention. Synchronized speech-and-action systems narrow the interval between interpretation and a prepared call \cite{PAPER_09240145263B}. They make it especially important to distinguish a candidate operation from one authorized for execution, and to invalidate decisions that depend on superseded speech.

\subsection{Memory and Personalization}
\label{sec:agentic-memory}

Memory allows a voice agent to use information beyond the current utterance, although several mechanisms are often discussed under the same name. \emph{Interaction context} retains the recent exchange needed to interpret a turn. \emph{Retrieval memory} supplies information from an external store, including through audio-aware or transcription-free retrieval \cite{PAPER_76F78E88F459,PAPER_7C61510F1269}. \emph{Persistent adaptive memory} carries preferences, observations, or episodes across sessions and associates them with an identity \cite{PAPER_FFAF306C931D,PAPER_07DE94B0DB13}. The last category creates a particular responsibility: information gathered in one conversation may influence an action long after the circumstances in which it was recorded have changed.

Retrieval has an established methodological foundation. Dense Passage Retrieval learns passage retrieval for open-domain question answering~\cite{karpukhin2020dense}, while RETRO demonstrates language modeling supported by an external token database~\cite{borgeaud2022improving}. Voice-oriented systems may keep knowledge outside the speech model so that it can be updated independently \cite{PAPER_45B6406FAE17}, retrieve asynchronously without blocking an ongoing exchange \cite{PAPER_71E72FFC5BF2,PAPER_F42079EE4844}, or use acoustic evidence rather than a single transcription hypothesis \cite{PAPER_76F78E88F459}. Recall, answer accuracy, and faithfulness describe how well information is retrieved and used to answer a question. They do not establish whether a retrieved preference belongs to the current speaker, remains valid, or may legitimately inform an action.

Persistent memory is better examined through the lifecycle of an entry than through retrieval alone. A useful set of operations is \emph{write $\rightarrow$ retrieve $\rightarrow$ update $\rightarrow$ consolidate $\rightarrow$ correct/delete}. Writing determines what becomes durable; retrieval brings earlier information into a decision; updating and consolidation reconcile new observations with what is already stored. Correction and deletion allow users to revise or withdraw information that would otherwise continue to shape behavior. These operations also need identity boundaries on shared devices \cite{PAPER_4AC64B8C94AB}. Enrollment, speaker verification, and identification are distinct tasks, and recognizing a voice does not itself establish permission to access or act on associated memory \cite{PAPER_E648C7BE210D}.

Two failures become particularly consequential when memory influences planning. \emph{Temporal drift} concerns deterioration in inference, memory, or conversational coherence as audio context accumulates, limiting the conclusions that can be drawn from short sessions \cite{PAPER_AA174144A864}. \emph{Provenance loss} occurs when summarization or consolidation obscures the source, time, or subsequent correction of a stored statement. Either can change a plan or tool argument without leaving an intelligible account of why it changed.

For consequential use, a memory entry should therefore retain its identity scope, source, temporal context, and revision history. On-device and privacy-oriented approaches can reduce some exposure of personal information \cite{PAPER_E5C507BB2EE0,PAPER_49C7AB780EBD}, while agent frameworks can expose lifecycle events to other components \cite{PAPER_40A1EF47A5D0}. The difficult part is carrying a correction beyond the memory store, to any pending decision that relied on the obsolete information.

\subsection{Retrieval and Tool-Augmented Voice Agents}
\label{sec:agentic-tools}

Tools provide a route from spoken requests to external services. Selecting a function and generating its arguments are well-established research problems \cite{schick2023toolformer,patil2024gorilla}. One dataset contains thirty thousand voice-cloned and noisy requests from smart-car, smart-home, and wearable domains, with single-intent, multi-intent, and needle-in-a-haystack settings; both acoustic perturbation and compositional demands reduce selection performance \cite{PAPER_766B59A1727C}. Real-time evaluations add disfluent human speech and report tool-selection $F_1$ and argument accuracy \cite{PAPER_9AD353416653}. Other frameworks check whether generated calls match their specified targets \cite{PAPER_2423A4A10719,PAPER_B51A1562076B,PAPER_D8B03F9E8908}.

These tests establish call formation, not necessarily authorization, execution, or success. The distinction is particularly important when action generation and speech share a timeline. Duplex models can produce planning text, JSON calls, and turn-control labels while the user is still speaking \cite{PAPER_09240145263B}; self-correction tests examine whether an earlier intent is abandoned after the speaker revises it \cite{PAPER_9AD353416653}. A call may therefore be ready before the utterance that justified it has been fully resolved. A correct final reply does not establish that the prepared operation was safely withheld or revised.

Grounded environments provide stronger evidence by checking the resulting task state. A benchmark of 278 retail, airline, and telecom tasks combines explicit policies, callable tools, and a target database state. For audio-native providers, it reports 31--51\% clean and 26--38\% realistic pass@1, with a substantial portion of voice- and noise-sensitive failures attributed to agent behavior rather than the acoustic front end alone \cite{PAPER_535B4C25EDC0}. Agent frameworks also expose typed executors, safety components, and lifecycle events; one reported planner path meets its working criterion across thirteen request classes, with weaker fallback behavior in other configurations \cite{PAPER_40A1EF47A5D0,PAPER_B421DE5BB9A8}. Applications range from rural form filling and conversational search to domain support and multimodal skill invocation \cite{PAPER_3FD26CDF1D8B,PAPER_838318340492,PAPER_0BB1928924AB,PAPER_E4FD61E964C3}, including clinical scheduling and prehospital assessment \cite{PAPER_016CD07D8596,PAPER_6688E5A9A34D}. Their reported outcomes must be interpreted in light of the particular environment and task definition.

Commercial accounts describe a further step toward operational use. Yuma Voice AI documents workflows in which an agent identifies a caller, retrieves order information, obtains verbal confirmation, and performs backend operations such as address changes, cancellations, refunds, or subscription updates. Transcripts and configurable human handoff are also described \cite{YUMA_VOICE_2026,YUMA_VOICE_LAUNCH_2026}. This is evidence of a vendor-reported production workflow linking speech to services with side effects, not an independent evaluation of authorization accuracy, post-action verification, rollback, or long-term user outcomes.

Even with a typed executor, the scope of available operations constrains what a system can attempt, and a well-formed call may still be unauthorized or rejected \cite{PAPER_40A1EF47A5D0}. Real services add timeouts, malformed responses, concurrent changes, and partial failures that deterministic mock tools do not fully capture \cite{PAPER_9AD353416653}. Evaluating tool use therefore requires a distinction between producing a call, dispatching it, and establishing what happened afterward.

\subsection{Digital and Physical Action}
\label{sec:agentic-embodied}

Physical action makes the consequences of spoken interpretation harder to ignore. PaLM-E combines embodied observations with language modeling~\cite{driess2023palm}; RT-1 learns robot control from real-world data~\cite{brohan2022rt}; and RT-2 connects vision-language representations to robotic actions~\cite{brohan2023rt}. Voice-oriented systems add the timing and ambiguity of conversation. DOROTHIE links dialogue to navigation in simulation \cite{PAPER_A40BBDA83DC5}; other systems use multimodal context for proactive manipulation or low-latency service-robot interaction \cite{PAPER_E9854767F763,PAPER_4C2C38D58510}. Egocentric assistants must resolve references against the wearer's view \cite{PAPER_A8347D87D2F8}, and spoken human--robot collaboration depends on establishing shared context \cite{PAPER_CBCDEABE5D3A,PAPER_295CA1FD8302}.

Concurrent speaking and acting can introduce competition between speech and action processing, with one reported system showing lower long-horizon task performance when the two operate together \cite{EXTERNAL_ELLSA}. MIRA addresses a related timing problem through turn arbitration, barge-in handling, and streaming motion generation. Its controller releases only a short committed motion prefix and applies robot-side joint and collision checks, leaving later predicted motion open to revision \cite{MIRA_2026}. This bounds how much physical behavior is committed while spoken evidence remains provisional, although it does not establish independent outcome verification or compensation after failure.

In embodied settings, several familiar properties of spoken interaction become control requirements.

\textbf{Reference grounding.} An instruction such as ``the one on the left'' refers to a scene that may change before the robot acts. Situated systems must therefore maintain and, when necessary, revise the referent \cite{PAPER_A40BBDA83DC5,PAPER_A8347D87D2F8}.

\textbf{Timing.} Speech arrives while control is ongoing. A stop command may need to interrupt motion before its current trajectory finishes, making endpointing and duplex control relevant to safety as well as responsiveness \cite{PAPER_4C2C38D58510}.

\textbf{Speaker identity.} In shared spaces, recognizing who spoke is not sufficient to determine who may issue a particular command. The attribution mechanisms discussed in Section~\ref{sec:interaction} must be connected to the authorization policy.

\textbf{Correction and confirmation.} A revision may arrive after movement has begun. Unlike a candidate plan, a physical change cannot necessarily be undone by editing a record; stopping, observing the result, and compensating where possible become part of the control problem \cite{EXTERNAL_ELLSA}.

\textbf{Human takeover.} A robot operating near people needs a clear means of returning control. Studies of attention and listening in assistive settings examine takeover prompts and hand-back behavior \cite{PAPER_8554E35C4B44}.

\textbf{Recovery during action.} A failed grasp, an unexpected obstacle, or an interrupted instruction may require a response before the next conversational turn. Recovery must therefore be evaluated while control is still active.

Evidence for these requirements comes from different experimental settings. Habitat provides a platform for embodied simulation~\cite{savva2019habitat}, while ALFRED evaluates language-guided household tasks~\cite{shridhar2020alfred}. Driving simulators report navigation outcomes under controlled interaction \cite{PAPER_A40BBDA83DC5}; laboratory studies examine naturalness, appropriateness, and takeover over relatively short sessions \cite{PAPER_8554E35C4B44}; and supervised workflows provide narrower deployment evidence \cite{PAPER_6688E5A9A34D}. Simulation can model selected disturbances without reproducing the full range of acoustic, perceptual, and operational conditions encountered in deployment. Short laboratory studies also offer limited evidence about persistent state over time \cite{PAPER_A8347D87D2F8,PAPER_9DF653778FD9}.

Other work connects speech to visual context \cite{PAPER_DB98248B80C7,PAPER_020E54A469DD,PAPER_E329E5863B5F}, brings reasoning and tools into spoken dialogue \cite{PAPER_2A446EEB5470,PAPER_AC489EF2815D}, and evaluates limit awareness or multi-step assistance rather than response quality alone \cite{PAPER_CDE5AB97CCAE,PAPER_B1FA1724AA27,PAPER_97F170FBE732}. These advances make it increasingly important to observe the effects of an action, not simply the agent's ability to propose one.

\subsection{Action Closure: From Invocation to Verified Consequence}
\label{sec:agentic-ladder}

A valid tool call does not establish that the intended change occurred. We use \emph{action closure} for the consequence-side loop \emph{execute $\rightarrow$ observe $\rightarrow$ verify $\rightarrow$ recover or compensate}. This is part of the broader agency closure introduced in Section~\ref{sec:intro-thesis}, which also includes spoken interaction, authorized commitment, and communication of the verified outcome to the user.

The stages address different questions. \emph{Execution} concerns whether an operation was dispatched and what the service or controller did with it. \emph{Observation} records the state reported after the attempt. \emph{Verification} compares that state with the intended result. If they differ, \emph{recovery} may require retrying, replanning, compensation, or escalation. A selected function may never be invoked; an accepted request may produce only a partial change; and a reported state may not satisfy the user's goal. A single success label obscures these distinctions and gives little guidance about where intervention is needed.

\begin{figure*}[t]
    \centering
    \includegraphics[width=\textwidth]{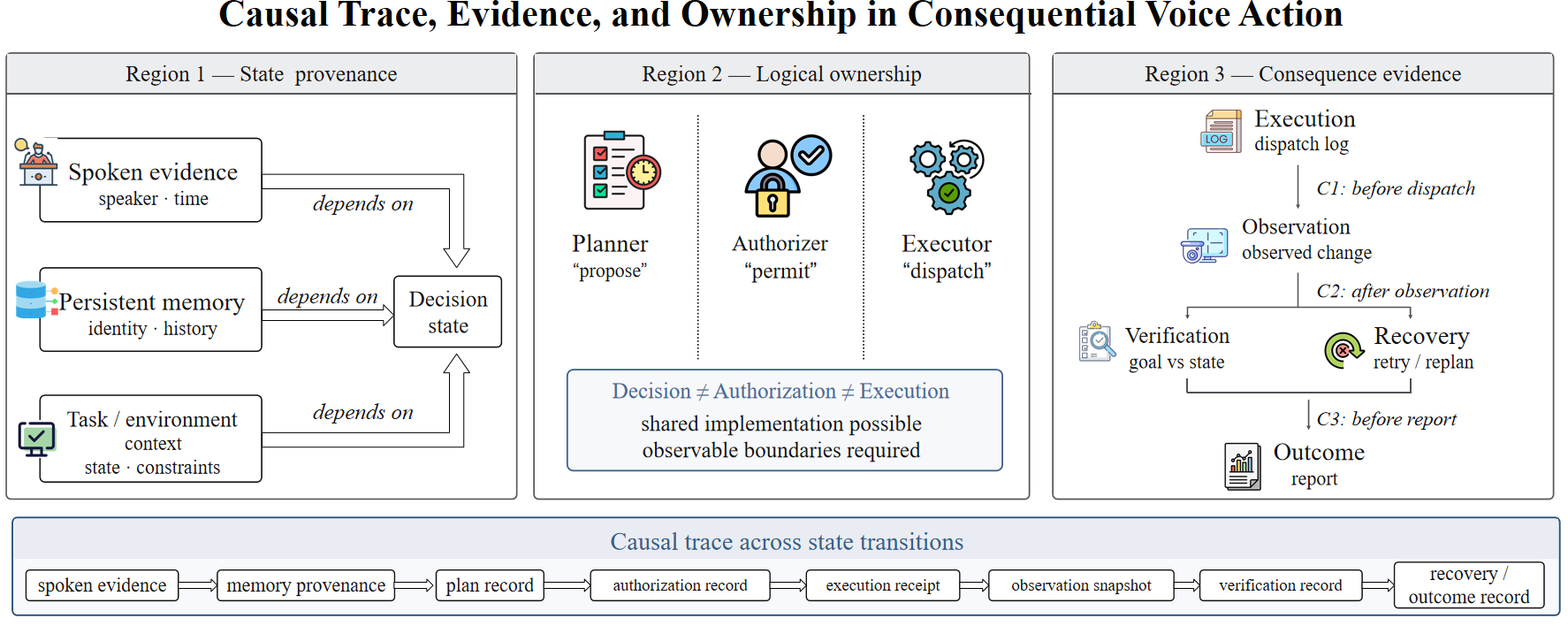}
    \caption{Causal trace, evidence, and ownership in consequential voice action. Spoken evidence, persistent memory, and task context inform decision state. Planning, authorization, and execution have distinct logical roles, while observation, verification, and recovery provide evidence of the outcome. The bottom band traces these dependencies; C1--C3 mark opportunities to contain errors before dispatch, after observation, and before success is reported.}
    \label{fig:agent-loop}
\end{figure*}

Figure~\ref{fig:agent-loop} presents the loop as a map of responsibility and evidence. Spoken input, memory, and environmental context inform a decision, but proposing, authorizing, and executing an operation remain logically distinct even when they share an implementation. Three points are particularly useful for containing errors: before dispatch, after the result is observed, and before success is communicated.

Existing evaluations illuminate different parts of this loop. Speech-to-function and duplex benchmarks test function selection and argument generation under noise, disfluency, or interruption \cite{PAPER_766B59A1727C,PAPER_9AD353416653}; agent frameworks expose typed execution and structured failure handling \cite{PAPER_40A1EF47A5D0,PAPER_B421DE5BB9A8}; grounded task benchmarks compare a resulting database state with a specified goal \cite{PAPER_535B4C25EDC0}; and interruption studies distinguish recovery quality from task completion \cite{PAPER_2D9099D449E2}. Each supplies evidence at a particular point, but none of the call-level measures alone establishes a verified consequence.

These are logical responsibilities, not necessarily separate software modules. A planner proposes an operation, an authorization mechanism determines whether it may proceed, an executor carries it out, and verification assesses the observed result. One model or runtime may implement several roles, provided the relevant transitions and evidence remain accessible. Otherwise, a fluent spoken confirmation may express what the system expected to happen rather than what it has established actually happened.

The same boundaries also determine when human control becomes necessary. An agent that can change persistent or external state must establish who may authorize the change, which information it may use, and how a user can intervene when the result is wrong or uncertain. These questions motivate the governance and human-control mechanisms discussed next.

\section{Human-Centered and Trustworthy Voice Agents}
\label{sec:human}

Trustworthiness is part of the same state transitions that make a voice agent consequential. Identity determines whose evidence enters the loop; authorization limits which plans may be executed; privacy and consent constrain persistent state; accessibility determines who can use the interaction reliably; and verification and recovery bound the effect of an incorrect action. These are architectural constraints as well as deployment policies. The ownership and evidence boundaries summarized in Fig.~\ref{fig:agent-loop} provide a common frame for asking how identity, permission, privacy, accessibility, and oversight constrain consequential state transitions.

\subsection{Personalization, Identity, and Authorization}
\label{sec:human-identity}

Personalization introduces durable state into the loop, and durable state brings obligations with it. Writing a preference, retrieving it later, and acting on it are three separate decisions. Category-bounded memory controls unbounded growth by organizing long-term information about a user \cite{PAPER_FFAF306C931D}, and streaming dual-store memory makes retrieval available during an exchange rather than between turns \cite{PAPER_07DE94B0DB13}. Neither design answers the question a user is entitled to ask: what is stored about me, why, and how do I remove it. A preference hit-rate metric is a measure of usefulness and not of control.

Identity is a distinct technology. The underlying studies separate enrollment, verification, and identification as distinct tasks \cite{PAPER_E648C7BE210D}. This separation is what makes authorization expressible. A system may recognize a voice and still be required to establish that the person is entitled to issue a particular instruction, and identity-aware memory shows that the failure mode is concrete: on a shared device, context can leak across speakers unless isolation is designed in \cite{PAPER_4AC64B8C94AB}. Recognition is evidence about who spoke; it is not authorization to act. Spoofing countermeasures are also relevant to the reliability of identity evidence: ASVspoof~2019 evaluates the detection of spoofed and fake audio~\cite{todisco2019asvspoof}.

Authorization is the property that converts a technically valid action into a permitted one, and it constrains system design in a specific way: the component that decides and the component that acts must be separated by a boundary that carries identity and scope. Governance frameworks for tool-using voice assistants have begun to specify integrity, access control, and audit requirements, and they demonstrate that restrictions change measured leakage \cite{PAPER_BA00BDF2544F}. Available evaluations do not yet combine denied, stale, ambiguous, and revoked permissions in a spoken interaction while measuring both unsafe actions and correct abstentions. In their absence, tool-selection accuracy and pass rates cannot substitute for an authorization measurement.

\subsection{Consent, Privacy, and Security}
\label{sec:human-privacy}

Consent in a voice agent is not a single event at enrollment but a lifecycle that spans capture, retention, reuse, and deletion. Privacy-preserving on-device assistants show that some of the associated risk can be reduced by keeping audio local \cite{PAPER_E5C507BB2EE0}, and interactional privacy benchmarks measure what a speech model discloses about a conversation rather than only what it transcribes \cite{PAPER_49C7AB780EBD}. Both are measurements of an interface:
the first of where computation happens, the second of what the model reveals. A distinct privacy risk arises from model memorization: training-data extraction studies show that language models can reveal portions of their training data~\cite{carlini2021extracting}.

Security results are especially relevant to architecture because persistence and action enlarge the attack surface. Earlier attacks on speech-to-text systems demonstrated targeted adversarial transcription~\cite{carlini2018audio} and the hiding of perturbations using psychoacoustic constraints~\cite{schonherr2018adversarial}. Speech-specific adversarial inputs can hijack audio-language models through imperceptible perturbations \cite{PAPER_AF925F6B6A7D}, and defenses have been proposed that patch speech language models against jailbreak attempts \cite{PAPER_FFE347059994}. Broader audio safety tooling spans multiple threat models \cite{PAPER_8791531A31B8}.
Persistent, action-capable agents expose two additional failure modes. Speaker inversion shows that speech tokens can leak identity-relevant information even when transcripts appear harmless \cite{PAPER_1036315221F1}, and denial-of-service attacks show that an agent can be kept talking, and therefore kept from acting, by adversarial input \cite{PAPER_E62762C38D81}. A system that remembers and can call tools can be attacked through its memory and its tools, not only through its recognizer~\cite{greshake2023not}, and the corresponding design consequence is that authorization, memory writes, and tool invocations each require their own policy and their own audit record.

Measurement infrastructure is developing in parallel. Manipulation detection in multi-speaker speech, counterfactual audits of whether paralinguistic evidence is actually used, and studies of judge reliability with human oversight all address the question of whether a claimed capability is the capability that was measured \cite{PAPER_93B59339D43D,PAPER_AB7D7F2FBF29,PAPER_58D5F87AB365}.

\subsection{Trust, Empathy, and Interaction Quality}
\label{sec:human-trust}

Human-facing evidence is concentrated in attentive listening and embodied conversation. Studies with older adults report how often responses were judged appropriate, with a substantial remainder judged inappropriate, which makes the limits of the behavior visible rather than aggregated away
\cite{PAPER_12AD80B5D1F4}. Expressive virtual coaching systems document their architecture and evaluation in stages \cite{PAPER_307D36780825,PAPER_AAD7A364FA60},
affect-annotated corpora provide conversational data for complaint handling and listening \cite{PAPER_C2C0128ACA13}, and recent systems push empathy into the dialogue model itself, including full-duplex empathic models and empathy training with frozen language models \cite{PAPER_BEE9E2EA0BDC,PAPER_170D457B6045}.

Timing, backchannels, prosody, and embodiment change how users interpret an agent and can affect what users are willing to say or authorize. The evidence should not be stretched beyond those endpoints. Perceived empathy is not evidence of emotion understanding, expressive output does not establish safe handling, and appropriateness ratings from short sessions do not predict long-term relational outcomes. Trust therefore needs its own measurements, including measurements over time~\cite{lee2004trust,hoff2015trust}.

Usability evidence follows the same pattern. Studies of voice agents inside existing workflows report usability scores without integrating a real record system or establishing outcomes \cite{PAPER_D0815A91F238}; field evaluations of conversational voice diaries measure adherence and richness of self-report \cite{PAPER_298143D3B9A0}; studies of older adults' perception of assistant explanations show that personality and explanation style change acceptance \cite{PAPER_3389FD1BA564}; and classroom studies show that perceived accent shapes how users collaborate with a voice system \cite{PAPER_F5E043F38380}. Usability is a legitimate endpoint when the system is a tool, and it becomes a proxy error when it is used to argue for safety: a high usability score establishes that people could use the system, not that the system made good decisions.

\subsection{Accessibility, Fairness, and High-Stakes Use}
\label{sec:human-fairness}

Equity evidence appears as language, accent, age, and disability coverage, and it matters technically because recognition and synthesis quality determine whether a person can use the agent at all. Documented racial disparities in automated speech recognition provide a concrete example of how recognition performance can differ systematically across user groups~\cite{koenecke2020racial}. Multi-accent benchmarks, code-switched evaluations, accent-robust recognition, and conversational systems for low-resource varieties show who is at risk of exclusion
\cite{PAPER_A8FB29F113AF,PAPER_3FE31C58A49B,PAPER_5C4380166688,PAPER_F2778FA85677}.
Domain-specific recognition failures concentrate in exactly the settings where errors are consequential: code-switched clinical speech produces substantial task-critical word error even when overall recognition appears acceptable \cite{PAPER_640D8E073990}. Acceptance also varies by population independently of underlying recognition performance \cite{PAPER_3389FD1BA564,PAPER_F5E043F38380}. Intersectional and longitudinal equity evidence is limited in the studies considered here. The evidence summarized above provides language-, accent-, age-, and setting-specific slices, but little basis for conclusions about their intersections or about outcomes over time.

High-stakes use makes the distinction between assistance and authority concrete. Structured prehospital assessment uses a voice agent inside a clinical workflow with clinician confirmation \cite{PAPER_6688E5A9A34D}; prescription and consultation systems adapt speech models to medical language \cite{PAPER_E24D573FC145}; and reviews map where voice and language interventions appear in mental-health care \cite{PAPER_BD0C791C1040}. Consent and ethics of collecting multimodal conversation with people who have cognitive impairments have been treated explicitly as a research design problem \cite{PAPER_AD8CAF8EB889}. In these settings the relevant endpoints are clinical confirmation, escalation, false-negative rates, and professional oversight, none of which is implied by conversational quality, and the health-related results considered here are best read as evidence about workflow fit rather than medical efficacy.

\subsection{Governance as a Property of the Loop}
\label{sec:human-governance}

A persistent, action-capable agent couples these governance problems mechanically. A lifecycle perspective on machine-learning harms helps distinguish risks originating in data, model development, evaluation, and deployment~\cite{suresh2021framework}. Wrong speaker attribution becomes a privacy breach when the system discloses something to the wrong person. A misunderstood correction becomes an unauthorized action when a call has already been prepared. A persuasive voice amplifies consent problems when the system is asking for permission. A memory that cannot be deleted undermines the assurance given at enrollment. An irreversible physical action removes the recovery path that confirmation is supposed to provide.

At the ownership and evidence boundaries summarized in Fig.~\ref{fig:agent-loop}, the system must therefore answer concrete questions: who spoke; was the utterance addressed to the
agent; was the speaker authorized; did the user consent; what action was
actually executed; what evidence establishes the resulting state; can it be reversed; and what does the agent tell the user
afterwards. These conditions determine whether the commitment and consequence stages of agency closure are auditable: an action that cannot be attributed to a speaker, authorized against
an identity, or explained afterwards has closed the loop in software terms and not in human terms. Table~\ref{tab:trustworthy} places these controls beside the capability interfaces they constrain. The juxtaposition is deliberate: evidence that a system can remember, plan, or execute does not establish that identity, consent, authorization, or escalation was handled correctly. The evaluation section therefore treats these controls as observable requirements for consequential systems.
% Table III: consolidated agentic capability + governance landscape.

\begin{table*}[tp]
\centering
\caption{Agentic capabilities and governance constraints organized by the interface they expose. Capability evidence and human-centered constraints are shown together because consequential behavior depends on both what the system can do and the conditions under which that transition is permitted and observable.}
\label{tab:agentic-governance}
\label{tab:reasoning-memory-tools}
\label{tab:trustworthy}

\begingroup
\scriptsize
\setlength{\tabcolsep}{5pt}
\renewcommand{\arraystretch}{1.12}
\setlength{\extrarowheight}{1pt}
\setlength{\aboverulesep}{0pt}
\setlength{\belowrulesep}{0pt}
\arrayrulecolor{RoadRule}

\begin{tabularx}{\textwidth}{
  R{2.70cm} Z{0.95} Z{0.85} Z{1.20}
}
\toprule[0.8pt]
\rowcolor{RoadHead}
\textcolor{RoadInk}{\textbf{Interface}} &
\textcolor{RoadInk}{\textbf{Mechanism or system question}} &
\textcolor{RoadInk}{\textbf{Useful endpoint / horizon}} &
\textcolor{RoadInk}{\textbf{Boundary and representative evidence}} \\
\midrule[0.5pt]

% ---------- A. State, reasoning, and memory ----------
\rowcolor{RoadHead!55!white}
\multicolumn{4}{l}{
  \textcolor{RoadInk}{\textbf{A. State, reasoning, and memory}}
} \\[3pt]
\midrule[0.3pt]

\textbf{Explicit state / policy}
&
symbolic or learned dialogue state with a policy over declared variables
&
state accuracy; policy return; task success / turn--session
&
makes decision state inspectable
\cite{PAPER_E17F117D3387,PAPER_03CE3B0B842C,PAPER_EC2365EC8FAE}
\\[2pt]
\midrule[0.3pt]

\textbf{Streaming reasoning}
&
partial transcript, semantic event, or latent speech state triggers deliberation
&
trigger quality; reasoning quality; latency / turn--session
&
reasoning can start before a turn is complete, but early hypotheses remain provisional
\cite{PAPER_13438B5C64C7,PAPER_CB6C6E4E2062,PAPER_8B3C1EFAB4CB}
\\[2pt]
\midrule[0.3pt]

\textbf{Clarification / replanning}
&
uncertainty-aware question, revision, or abandonment of a provisional plan
&
clarification accuracy; revised task outcome / task
&
distinguishes revising a plan from merely producing another response
\cite{PAPER_EC0445D89B24,PAPER_913915B00C96}
\\[2pt]
\midrule[0.3pt]

\textbf{Retrieval memory}
&
audio- or text-based retrieval into the current interaction
&
recall; answer accuracy; faithfulness / turn--session
&
retrieval quality does not establish durable user state
\cite{PAPER_76F78E88F459,PAPER_7C61510F1269,PAPER_45B6406FAE17}
\\[2pt]
\midrule[0.3pt]

\textbf{Persistent memory}
&
preference, episodic, or dual-store memory across sessions
&
preference use; identity isolation; long-horizon quality / cross-session
&
persistence creates correction, deletion, provenance, and drift obligations
\cite{PAPER_FFAF306C931D,PAPER_07DE94B0DB13,PAPER_4AC64B8C94AB}
\\[3pt]
\midrule[0.5pt]

% ---------- B. Action and consequence ----------
\rowcolor{RoadHead!55!white}
\multicolumn{4}{l}{
  \textcolor{RoadInk}{\textbf{B. Action and consequence}}
} \\[3pt]
\midrule[0.3pt]

\textbf{Speech-to-tool mapping}
&
function selection and structured argument generation from speech
&
selection and argument accuracy / turn
&
valid call generation is not execution
\cite{PAPER_766B59A1727C}
\\[2pt]
\midrule[0.3pt]

\textbf{Synchronized tool use}
&
action tokens or structured calls generated on the duplex speech clock
&
tool accuracy; pass@1; latency / turn--task
&
timing evidence remains distinct from authorization and side effects
\cite{PAPER_09240145263B,PAPER_9AD353416653}
\\[2pt]
\midrule[0.3pt]

\textbf{Typed execution / recovery}
&
planner dispatches constrained operations with retry or fallback
&
completion; lifecycle events; fallback behavior / task
&
execution is observable, but success still requires outcome verification
\cite{PAPER_40A1EF47A5D0,PAPER_B421DE5BB9A8}
\\[2pt]
\midrule[0.3pt]

\textbf{Grounded state change}
&
workflow evaluated against a target environment or database state
&
achieved-vs.-target state / task
&
establishes consequence inside the stated environment
\cite{PAPER_535B4C25EDC0}
\\[2pt]
\midrule[0.3pt]

\textbf{Embodied action}
&
dialogue state coupled to navigation, manipulation, or situated control
&
task outcome; control success; takeover / episode
&
physical consequence raises reversibility and hand-back requirements
\cite{PAPER_A40BBDA83DC5,PAPER_E9854767F763,EXTERNAL_ELLSA}
\\[3pt]
\midrule[0.5pt]

% ---------- C. Human control, security, and deployment ----------
\rowcolor{RoadHead!55!white}
\multicolumn{4}{l}{
  \textcolor{RoadInk}{\textbf{C. Human control, security, and deployment}}
} \\[3pt]
\midrule[0.3pt]

\textbf{Personalization / memory control}
&
What user context may persist, and how can it be corrected or removed?
&
preference use; correction; deletion; drift
&
useful memory is not equivalent to controllable memory
\cite{PAPER_FFAF306C931D,PAPER_07DE94B0DB13,PAPER_4AC64B8C94AB}
\\[2pt]
\midrule[0.3pt]

\textbf{Identity / authorization}
&
Who is speaking, who is being served, and may this speaker request this action?
&
verification; unsafe-action rate; correct abstention
&
recognition is evidence of identity, not permission to act
\cite{PAPER_E648C7BE210D,PAPER_BA00BDF2544F,PAPER_40A1EF47A5D0}
\\[2pt]
\midrule[0.3pt]

\textbf{Consent / privacy}
&
What may be captured, retained, reused, or disclosed?
&
consent lifecycle; disclosure; reconstruction risk
&
privacy constraints travel with state, not only with the microphone
\cite{PAPER_E5C507BB2EE0,PAPER_49C7AB780EBD,PAPER_1036315221F1}
\\[2pt]
\midrule[0.3pt]

\textbf{Security / availability}
&
Can adversarial audio, memory, or tool inputs alter or block the loop?
&
attack success; defence rate; leakage; availability
&
persistence and action enlarge the attack surface
\cite{PAPER_AF925F6B6A7D,PAPER_FFE347059994,PAPER_E62762C38D81}
\\[2pt]
\midrule[0.3pt]

\textbf{Trust / interaction quality}
&
Do timing, prosody, explanation, and recovery support calibrated reliance?
&
appropriateness; repeated-interaction reliance
&
perceived quality cannot substitute for safe or correct action
\cite{PAPER_12AD80B5D1F4,PAPER_307D36780825,PAPER_170D457B6045}
\\[2pt]
\midrule[0.3pt]

\textbf{Accessibility / fairness}
&
Who experiences degraded recognition, synthesis, or interaction quality?
&
subgroup differences across accent, language, age, and disability
&
access to the loop is itself a systems constraint
\cite{PAPER_A8FB29F113AF,PAPER_F5E043F38380,PAPER_3389FD1BA564}
\\[2pt]
\midrule[0.3pt]

\textbf{High-stakes oversight}
&
Which decisions require confirmation, escalation, takeover, or professional judgment?
&
escalation; takeover success; confirmed outcome
&
workflow fit does not establish domain efficacy
\cite{PAPER_6688E5A9A34D,PAPER_D0815A91F238,PAPER_8554E35C4B44}
\\[3pt]

\bottomrule[0.8pt]
\end{tabularx}
\endgroup
\end{table*}

\section{Evaluation: Matching Evidence to Agency Claims}
\label{sec:evaluation}

Evaluation makes the distinction between auditory intelligence and auditory agency concrete. Recognition, synthesis, latency, turn-taking, retrieval, and task completion each have established measures, but success on these measures does not necessarily establish that a voice agent produced the intended consequence. A system may recognize a request correctly, infer the intended task, and generate a valid tool call, yet act on behalf of the wrong speaker, proceed before a correction arrives, modify the wrong state, or report success without checking the result. Evaluating auditory agency therefore requires evidence that connects spoken input and its interpretation to commitment, authorized execution, and the consequences subsequently observed.

\subsection{From Component Performance to Consequence}
\label{sec:evaluation-hierarchy}

\begin{figure*}[tp]
\centering
\includegraphics[width=\textwidth]{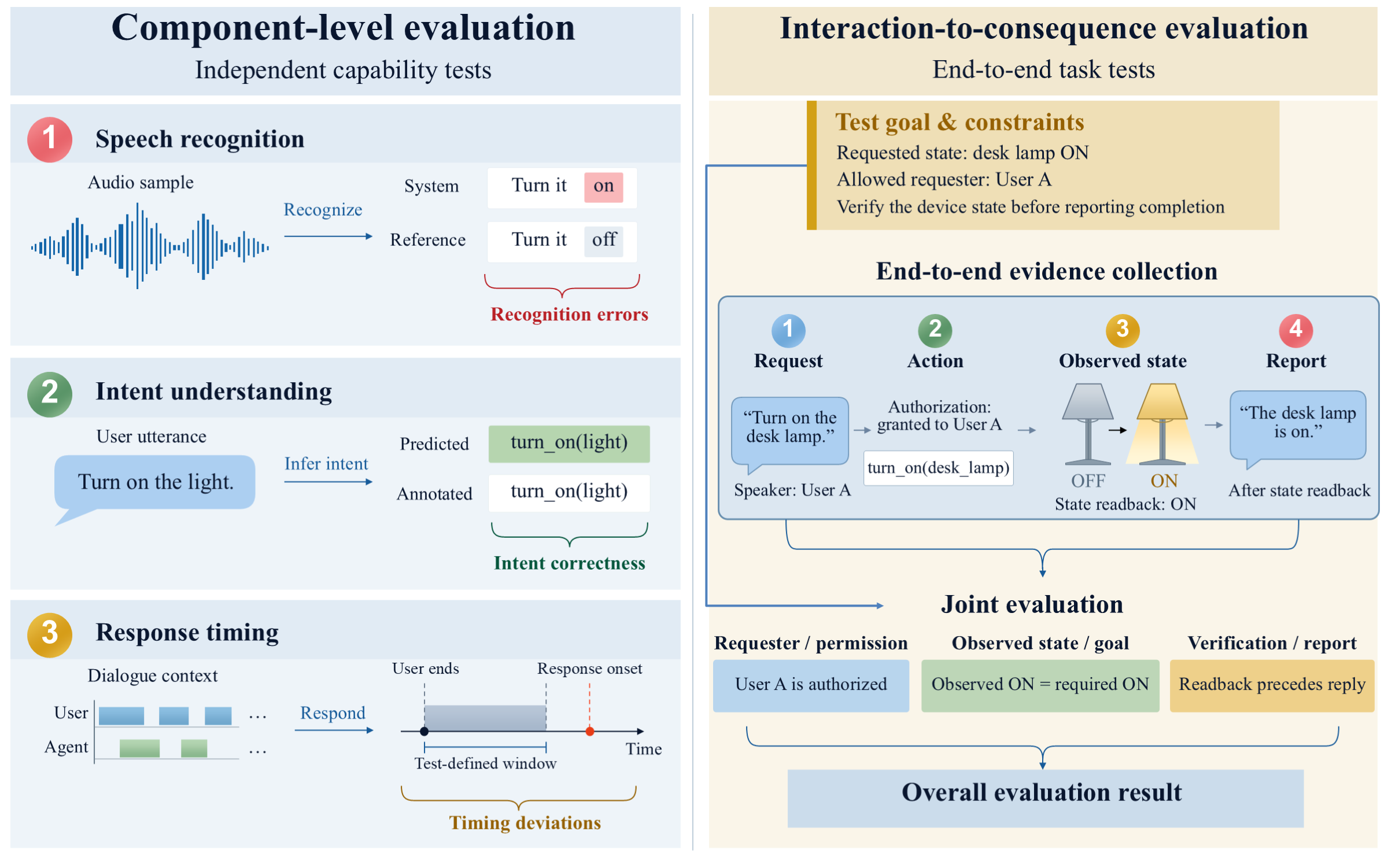}
\caption{Evaluation paradigms for voice agents. The figure contrasts the current component-level paradigm with the end-to-end interaction-to-consequence evaluations required by auditory agency, and summarizes the evaluation setup and metrics that make such claims testable.}
\label{fig:evaluation-hierarchy}
\end{figure*}

Speech systems have traditionally been assessed through instruments designed for particular capabilities. Figure~\ref{fig:evaluation-hierarchy} contrasts this component-centered practice with evaluations that follow an interaction through to its consequences. To clarify what different experiments can establish, we distinguish six recurring evidential endpoints.

At the \emph{signal} level, recognition, diarization, perceptual quality, and latency characterize the acoustic channel. The \emph{semantic} level concerns whether the system recovers the intended meaning, while the \emph{interaction} level examines timing, overlap, interruption, attribution, and repair. Together, these measurements describe how reliably a system can take part in spoken exchange. They do not, by themselves, establish what happens when the exchange leads to an action.

The remaining endpoints address that further step. A \emph{task} evaluation asks whether the system achieves a specified objective within a defined workflow or test environment; when a benchmark checks only a proposed tool call, its evidence is correspondingly limited to call formation. A \emph{consequence} evaluation examines the resulting external state and whether the intended change can be observed and verified. A \emph{human-outcome} evaluation considers how that behavior affects the person involved, including whether the system responds appropriately to correction, failure, or a request to regain control. These endpoints are not a universal ranking of systems. They describe the reach of the evidence supporting a particular claim.

The distinctions matter most at their boundaries. A correctly formed tool call may never be executed. A database benchmark may verify a final state within a simulator without establishing reliability in a live service. Even a verified state change can be inappropriate if the wrong person requested it or the system lacked authorization. Conversely, an endpointing study need not measure human outcomes to make a valid contribution. The point is to keep the conclusion within the scope of what was actually observed.

\subsection{What Existing Benchmarks Actually Expose}
\label{sec:evaluation-benchmarks}

Table~\ref{tab:results} brings together representative measurements across these endpoints. The results retain their original experimental settings and should not be interpreted as a common performance ranking.

% Table III in the compiled manuscript: representative evidence endpoints.

\begin{table*}[tp]
\centering
\caption{Representative evidence endpoints for voice agents. The endpoint determines the scope of the claim that can be supported. Reported values retain their original experimental meaning and are not directly comparable across rows.}
\label{tab:results}

\begingroup
\footnotesize
\setlength{\tabcolsep}{5pt}
\renewcommand{\arraystretch}{1.18}
\setlength{\extrarowheight}{2pt}
\setlength{\aboverulesep}{0pt}
\setlength{\belowrulesep}{0pt}
\arrayrulecolor{RoadRule}

\begin{tabularx}{\textwidth}{
  R{2.15cm} Z{1.00} R{2.65cm} Z{1.00}
}
\toprule[0.8pt]
\rowcolor{RoadHead}
\textcolor{RoadInk}{\textbf{Endpoint}} &
\textcolor{RoadInk}{\textbf{Representative evidence}} &
\textcolor{RoadInk}{\textbf{Reported result}} &
\textcolor{RoadInk}{\textbf{What the evidence establishes}} \\
\midrule[0.5pt]

\textbf{Signal}
&
four-state turn classification under the stated test conditions
\cite{PAPER_11D6F89D815C}
&
about 95.75\% mean accuracy
&
discrimination among pause, backchannel, interruption, and continuation
\\[5pt]
\midrule[0.3pt]

\textbf{Semantic}
&
streaming intent and slot localization in a closed-set setting
\cite{PAPER_C69F6F77FA7A}
&
about 98.97\% joint accuracy
&
meaning can be localized before a turn is complete
\\[5pt]
\midrule[0.3pt]

\textbf{Interaction}
&
barge-in prediction on real IVR audio
\cite{PAPER_C60EBC02D547}
&
macro-$F_1$ 0.826
&
when the system should yield; downstream state is not tested
\\[5pt]
\midrule[0.3pt]

\textbf{Task}
&
synchronized structured actions over 900 cases
\cite{PAPER_09240145263B}
&
85.56\% tool accuracy at 0.648\,s
&
low-latency call generation; execution and authorization remain separate
\\[5pt]
\midrule[0.3pt]

\textbf{Consequence}
&
278 grounded service tasks with a gold database state
\cite{PAPER_535B4C25EDC0}
&
31--51\% clean; 26--38\% realistic
&
achieved simulated state is checked against a target
\\[5pt]
\midrule[0.3pt]

\textbf{Human outcome}
&
attentive-listening interaction with older adults
\cite{PAPER_12AD80B5D1F4}
&
about 60\% judged appropriate
&
human-facing appropriateness over short interaction
\\[5pt]

\bottomrule[0.8pt]
\end{tabularx}
\endgroup
\end{table*}

The differences begin with what each benchmark treats as success. Streaming understanding evaluates intent and slot predictions before an utterance has ended \cite{PAPER_C69F6F77FA7A}. Turn-taking and barge-in benchmarks examine coordination under conversational timing \cite{PAPER_C60EBC02D547,PAPER_11D6F89D815C}, while duplex evaluations introduce overlap, disfluency, interruption, and self-correction \cite{PAPER_4F97D68B6C26,PAPER_9AD353416653}. Speech-to-tool benchmarks move closer to action by measuring the generation of structured calls \cite{PAPER_766B59A1727C,PAPER_09240145263B}. Grounded task environments can compare the resulting database state with a specified goal \cite{PAPER_535B4C25EDC0}, and recovery benchmarks examine behavior after an interruption or failure rather than treating ordinary task completion as sufficient \cite{PAPER_2D9099D449E2}. Each extends the observable part of the process, but none makes the other endpoints interchangeable.

Recent systems further illustrate why the word ``verification'' needs a precise object. CycleSpeech checks whether synthesized speech reflects intended vocal attributes. MIRA reports real-robot interruption handling and limits the portion of planned motion committed to execution. Yuma's product documentation describes live support calls that can trigger backend state changes \cite{CYCLESPEECH_2026,MIRA_2026,YUMA_VOICE_2026}. These are, respectively, evidence about speech generation, embodied control, and a reported production workflow. They should not be treated as equivalent demonstrations of independently verified agency closure; the last example, in particular, rests on vendor documentation rather than an independent evaluation.

Controlled benchmarks also leave out conditions that matter once a system acts. Supplying a transcript bypasses recognition uncertainty. Deterministic mock tools may omit denied requests, timeouts, partial writes, and concurrent state changes. Synthetic users may not reproduce the ways people revise instructions, withdraw consent, or correct an action already in preparation. Such simplifications are useful for isolating mechanisms. They become a limitation only when results obtained under those conditions are used to support broader deployment claims.

Time horizon creates a different boundary. Endpointing and tool-call tests often unfold over seconds or minutes, whereas persistent memory, identity isolation, and appropriate reliance emerge over repeated encounters. Studies of extended conversational audio have already reported degradation in inference, memory, and self-coherence as context accumulates \cite{PAPER_AA174144A864}. An evaluation of persistent agency must therefore last long enough to observe the state and behavior it claims to measure.

\subsection{Metrics Are Defined by Their Boundaries}
\label{sec:evaluation-reporting}

Even familiar metrics can become difficult to interpret when their event boundaries differ. ``Latency'' may refer to the delay of a model chunk, the time to first audio, the interval from an estimated endpoint to the first response, or the time from microphone input to a tool action. Synchronized action generation can report sub-second delays even when completion of an interactive task takes several seconds \cite{PAPER_09240145263B,PAPER_9AD353416653}. These measurements are not contradictory; they describe different intervals. Comparisons are meaningful only when both the starting event and the observed endpoint are specified.

Automatic judgments and aggregate success rates require similar care. Language-model judges make large-scale assessments of recovery, naturalness, and task quality feasible, but their agreement is itself part of the measurement. In reported recovery experiments, agreement between automated judges exceeded agreement between judges and human raters \cite{PAPER_2D9099D449E2}. Calibration and targeted human assessment are therefore important when conclusions depend on these scores. Likewise, a single task pass rate can conceal whether failure arose in recognition, attribution, commitment timing, execution, or recovery. Studies that trace failures to particular interfaces \cite{PAPER_535B4C25EDC0,PAPER_4F97D68B6C26} provide information that an aggregate score alone cannot supply.

\subsection{Closed-Loop Evaluation as an Architectural Principle}
\label{sec:evaluation-observation}

In many current systems, architectural integration extends beyond what the evaluation records. A voice agent may combine streaming speech, reasoning, memory, tools, and recovery, yet its components are assessed through separate tests. Duplex systems can produce structured actions during speech \cite{PAPER_09240145263B}; agent frameworks expose typed execution and lifecycle events \cite{PAPER_40A1EF47A5D0,PAPER_B421DE5BB9A8}; and grounded environments can check whether a target database state was reached \cite{PAPER_535B4C25EDC0}. What is harder to reconstruct from these results is the sequence of decisions and observations linking the capabilities together.

Closed-loop evaluation need not mean building one benchmark that exercises every capability. The required trace depends on the claim. For a consequential voice task, it may include the spoken request, speaker attribution, provisional interpretations and their revisions, the commitment event, authorization, the proposed and dispatched action, the observed post-action state, and the evidence used to verify the outcome. If verification fails, the trace should also show whether the system retried, compensated, or handed control back. Memory-dependent tasks additionally require the provenance of the stored information used in the decision; embodied tasks require observations of the physical environment and the control handoff. Different benchmarks can examine different portions of this process, provided that the transitions relevant to their claims remain inspectable. AgentDojo, for example, offers a security-focused evaluation environment for prompt-injection attacks and defenses in tool-using agents \cite{debenedetti2024agentdojo}.

This requirement has implications for architecture as well as measurement. Auditing depends on records of the decisions and state transitions under examination \cite{raji2020closing}. An opaque component that decides, acts, and announces success can still be evaluated against an independently observed outcome, but its internal failure may be difficult to localize. Keeping decision and verification responsibilities logically distinct helps, even when they are implemented within the same model, so long as the evidence supporting each remains accessible and independently checkable. A model's own assertion that an action succeeded is not a substitute for observing the resulting state.

Component-level evaluation remains indispensable. Speech quality matters because errors in the input can propagate into the plan; interaction quality matters because a mistimed commitment can preserve an interpretation the user has already revised; and task competence matters because even a properly authorized action may fail. Closed-loop evaluation connects these measurements when a claim crosses their boundaries. Once a voice agent changes external state, evidence of the decision is no longer enough: the evaluation must also account for what happened afterward.

\section{Toward General Auditory Agency}
\label{sec:roadmap}

The remaining challenges in auditory agency arise where separately developed capabilities meet. Streaming models work with revisable interpretations, memory carries earlier interactions into later decisions, and planners and tools turn decisions into persistent or external effects. A correction may arrive at any point. The system must preserve the user's intent, the authority behind an action, and evidence of its outcome across these transitions.

Four interface problems organize the agenda. Table~\ref{tab:roadmap} frames them as experiments rather than maturity levels.

% Table IV in the compiled manuscript: research agenda.

\begin{table*}[tp]
\centering
\caption{Research agenda for unresolved voice-agent interfaces. Each direction pairs related work with an open systems question, an illustrative experiment, and proposed evaluation measures.}
\label{tab:roadmap}

\begingroup
\footnotesize
\setlength{\tabcolsep}{5pt}
\renewcommand{\arraystretch}{1.18}
\setlength{\extrarowheight}{2pt}
\setlength{\aboverulesep}{0pt}
\setlength{\belowrulesep}{0pt}
\arrayrulecolor{RoadRule}

\begin{tabularx}{\textwidth}{
  R{2.70cm} Z{0.95} Z{1.15} Z{0.90}
}
\toprule[0.8pt]
\rowcolor{RoadHead}
\textcolor{RoadInk}{\textbf{Research direction}} &
\textcolor{RoadInk}{\textbf{Core systems question}} &
\textcolor{RoadInk}{\textbf{Illustrative experiment}} &
\textcolor{RoadInk}{\textbf{Key evaluation measures}} \\
\midrule[0.5pt]

\textbf{Revisable spoken evidence}~
\cite{PAPER_82CBE66BE086,PAPER_9AD353416653}
&
How can partial speech guide planning while keeping dependent state revisable?
&
Introduce natural corrections at different stages of provisional planning; trace their propagation through beliefs, plans, and pending actions.
&
Revision accuracy; propagation latency; stale-state action rate.
\\[5pt]
\midrule[0.3pt]

\textbf{Identity, authority, and persistent state}~\cite{PAPER_4AC64B8C94AB,PAPER_AA174144A864}
&
How should identity, permissions, provenance, and validity periods remain attached to persistent decision state?
&
Track users and memory objects across sessions; introduce expired permissions, stale records, corrections, and deletion requests; test cross-user isolation.
&
Correction and deletion coverage; unauthorized reuse and cross-user leakage rates.
\\[5pt]
\midrule[0.3pt]

\textbf{Closing the action loop}~\cite{PAPER_535B4C25EDC0,PAPER_2D9099D449E2}
&
How should planners, executors, and verifiers coordinate action, verification, and recovery under failure or concurrent change?
&
Inject denials, timeouts, partial writes, and concurrent updates; trace authorization, abstention, execution, independent state checks, and recovery through cancellation, compensation, or escalation.
&
Verified task success; unauthorized action rate; recovery time; residual side effects.
\\[5pt]
\midrule[0.3pt]

\textbf{Longitudinal\newline coordination}~\cite{PAPER_12AD80B5D1F4,PAPER_3389FD1BA564}
&
How should memory, recovery, and hand-back adapt as user needs and reliance evolve across sessions?
&
Run repeated-session studies across diverse accessibility needs; vary memory continuity, recovery support, and hand-back policies; track user intervention and task resumption.
&
Reliance calibration; intervention effort; task resumption success; accessibility gaps.
\\[5pt]

\bottomrule[0.8pt]
\end{tabularx}
\endgroup
\end{table*}

\subsection{Reasoning Under Revisable Spoken Evidence}
\label{sec:roadmap-coupling}

A responsive voice agent may begin processing before the speaker finishes. Semantic endpointing and anticipatory turn prediction show that partial speech supports timing decisions \cite{PAPER_CB8E8D2C7474,PAPER_82CBE66BE086}, while incremental systems can initiate reasoning from semantic events without waiting for a complete utterance \cite{PAPER_13438B5C64C7}. User studies provide complementary evidence about the effects of incremental processing on interaction~\cite{chiba2025investigating}. The harder case is a correction that arrives after reasoning has begun.

Provisional computation needs to remain distinct from committed state. Early hypotheses can guide retrieval, tool prefetching, or candidate plans without authorizing irreversible operations. Decisions that affect persistent state or external services need an evidential record and a way to revise dependent decisions. Delay-trained endpointers make the timing of early decisions explicit \cite{PAPER_9DC02CA93758}, staged reasoning can shift some computation away from the response-critical path \cite{PAPER_8B3C1EFAB4CB}, and self-correction scenarios test whether a system abandons an obsolete intent when the user changes a request \cite{PAPER_9AD353416653}. These studies do not necessarily trace revisions into external state.

The limit of speculation can be tested directly: a user begins a request, the agent prepares a plan, and a later phrase invalidates a constraint. Evaluation would trace what was revised or cancelled and whether an external action escaped before the correction took effect. This connects early reasoning's latency advantage to correctness.

\subsection{Identity, Authority, and Persistent State}
\label{sec:roadmap-state}

Identifying a speaker is only part of deciding whether to act on a spoken request. Speaker verification, diarization, device-directedness, and addressee detection provide different forms of evidence about identity and conversational intent \cite{PAPER_E648C7BE210D,PAPER_40CE2E5F8957,PAPER_B49F83A8B5BB}. Authorization requires another judgment: whether that person may request the particular operation under the present conditions. On shared devices, the same request can have different consequences depending on who speaks.

Memory extends this problem across time. Cross-session and dual-store approaches retain preferences and interaction histories \cite{PAPER_FFAF306C931D,PAPER_07DE94B0DB13}, and identity-aware designs seek to keep those records associated with the appropriate user \cite{PAPER_4AC64B8C94AB}. For an acting agent, records also need identity, time, source, and correction or deletion paths.

Treating memory as \emph{state with provenance} lets a planner assess whether a preference remains current and applicable to a proposed action. If a user says ``I no longer want that,'' the correction may need to invalidate derived preferences, pending plans, or future uses of the original record. Evidence of degraded memory and self-coherence over accumulated interaction \cite{PAPER_AA174144A864} reinforces the need to track how stored information changes, not simply whether it can be retrieved.

Stale preferences, summary drift, cross-user contamination, and incomplete deletion may only emerge over repeated encounters. Longitudinal experiments should therefore follow both users and individual memory records across sessions, examining how revisions propagate into later decisions and whether obsolete information continues to influence action.

\subsection{Closing the Action Loop}
\label{sec:roadmap-action}

Tool-oriented voice systems are beginning to expose more than a final spoken answer. Speech-to-function datasets measure call selection under noise and compositional demands \cite{PAPER_766B59A1727C}; duplex models can prepare structured calls while an utterance is still unfolding \cite{PAPER_09240145263B}; and grounded benchmarks compare a resulting database state with a specified goal \cite{PAPER_535B4C25EDC0}. These are distinct points along the action path. Connecting them requires a control process that keeps authorization, execution, observation, verification, and recovery distinguishable.

This does not require a fully modular architecture. A planner may propose an operation, but the permission to execute it is a different decision. An executor can report the response of a service without determining whether the user's objective was achieved; that judgment depends on comparing observed and intended states. Recovery, in turn, depends on what the comparison reveals. Their transitions must remain observable even when one model performs several roles.

Live services introduce failures that deterministic mock interfaces often exclude: changed permissions, stale records, timeouts, partial writes, concurrent updates, and operations that cannot be undone. Evaluation should therefore reward \emph{correct non-actions} alongside successful execution. Refusing an unauthorized request, cancelling a prepared operation after a correction, or escalating when an outcome cannot be verified may be the appropriate result. Compensation after a partial write tests a still stronger capability. Existing recovery benchmarks distinguish repair quality from ordinary task success \cite{PAPER_2D9099D449E2}; an important next step is to evaluate recovery against observed external state rather than conversational repair alone.

Recent systems illustrate different parts of this challenge. MIRA limits the amount of robot motion committed while speech remains interruptible; Yuma's product documentation describes backend changes during a live support call followed by spoken confirmation; and CycleSpeech verifies whether generated speech matches intended vocal attributes \cite{MIRA_2026,YUMA_VOICE_2026,CYCLESPEECH_2026}. The last of these concerns speech fidelity rather than external action, and product documentation is not equivalent to independent outcome evaluation. Meanwhile, multimodal model releases and on-device platforms such as Snapdragon Gen 6 bring more of the computation into shared models and local hardware \cite{MIMO_V25_RELEASE_2026,QUALCOMM_GEN6_2026}. These examples do not establish an independently evaluated trace from overlapping speech through authorization, execution, verified external state, recovery, and a human-confirmed outcome. Establishing that trace is the central experimental challenge for action closure.

\subsection{Longitudinal Human--Agent Coordination}
\label{sec:roadmap-longitudinal}

As voice agents retain information and undertake tasks, users must decide when to rely on them and when to intervene. That decision depends on the agent's timing, transparency, accessibility, and handling of errors as much as on the naturalness of its speech. Short human studies already report variation in appropriateness, usability, and acceptance \cite{PAPER_12AD80B5D1F4,PAPER_D0815A91F238,PAPER_3389FD1BA564}; accent and language differences also affect who can use the interface reliably \cite{PAPER_A8FB29F113AF,PAPER_640D8E073990}.

Evidence from longer periods and real settings is emerging, although it addresses different questions. Cog-TiPRO followed daily voice-assistant commands from fifteen older adults in their homes over eighteen months to study cognitive change~\cite{qi2025cog}. This establishes longitudinal observation of voice-assistant use, not the reliability of an agent's persistent memory. A natural field experiment in hiring also reports downstream employment outcomes from interviews conducted by a voice agent, including higher offer rates, job starts, and worker retention without lower measured productivity \cite{fruits2026ai}. It provides evidence of real-world use and human consequences, but does not by itself test cross-session memory correction or independently verified action recovery.

Longer-term coordination raises further questions. Repeated success may encourage appropriate reliance or excessive delegation; repeated failure may lead to caution or abandonment. Personalization can reduce friction while allowing outdated assumptions to persist unnoticed. These effects are difficult to infer from a single-session preference rating, even when that rating is collected from real users.

Future evaluations could follow the same users while recording the agent's relevant state transitions. Does a corrected preference remain corrected in later sessions? Does an unsuccessful action change subsequent behavior? Does the agent seek clarification when evidence is weak, and can the user regain control promptly? Such tests would link human behavior to memory, decision, and recovery mechanisms without equating human-likeness with quality.

\subsection{Outlook}
\label{sec:roadmap-outlook}

The common requirement across these directions is \emph{controllable coupling}. A system may reason before a turn ends, provided that provisional interpretations do not silently become irreversible decisions. Persistent memory is useful when its provenance and revision history remain available to later decisions. Tool use becomes trustworthy when authorization and verification are distinguishable from call generation, and adaptation must leave the user able to redirect or regain control. These are interface requirements, not a prescription for a particular architecture.

No single benchmark is likely to capture all of them. Component-level evaluations remain valuable for isolating timing, recognition, retrieval, and tool-use mechanisms. They can be complemented by experiments that preserve a causal trace across selected boundaries: a correction arriving after a plan is prepared, an action influenced by stale memory, or a physical operation followed by observation, verification, and hand-back. This would make agency closure measurable without requiring every study to reproduce a full deployment.

The aim is not to collapse auditory agency into a single maturity score, but to establish which transitions a system can handle reliably and what evidence supports its claims of successful action.

\section{Conclusion}
\label{sec:conclusion}

Voice agents increasingly combine real-time speech, reasoning, memory, tool use, embodied control, and human-centered safeguards. Once spoken evidence can change persistent or external state, a list of component capabilities no longer describes the behavior that matters. \emph{Auditory intelligence} concerns participation in spoken interaction; \emph{auditory agency} begins when that interaction is causally coupled to a consequential decision.

Two interfaces make the coupling concrete. The \emph{commitment boundary} separates revisable evidence from state or action on which downstream components may depend. \emph{Agency closure} connects an authorized action to observation, verification, recovery, and communication of its outcome. The concepts are architecture-agnostic; even a unified model must make it possible to ask who or what owns a decision, what evidence justified it, what state was actually changed, and how disagreement between intended and observed outcome is handled.

Evaluation should follow the same discipline. Component and task metrics remain indispensable, but their evidential scope should not expand simply because the system integrates more functions. A claim about consequential agency requires evidence that reaches the consequence; a deployment claim requires evidence that reaches the user and the relevant time horizon. Progress will depend on architectures and evaluations that preserve the causal trace from interaction to outcome, so that the system's action-not only its speech or reasoning-can be examined and held accountable.

\bibliographystyle{IEEEtran}
\bibliography{references}

\end{document}